\documentclass[letterpaper,english,aps, prl, represent, twocolumn, groupedaddress, superscriptaddress]{revtex4-2}
\usepackage[T1]{fontenc}
\usepackage{color}
\usepackage{amsmath}
\usepackage{amssymb}
\usepackage{graphicx}

\makeatletter

\pdfpageheight\paperheight
\pdfpagewidth\paperwidth

\providecommand{\tabularnewline}{\\}
\newcommand{\lyxdot}{.}

\usepackage[bookmarks=true,colorlinks,linkcolor=black,anchorcolor=green,citecolor=blue,urlcolor=blue]{hyperref}

\renewcommand{\fnum@figure}{FIG. \thefigure}
\renewcommand{\fnum@table}{TABLE. \thetable}

\makeatother

\usepackage{babel}
\begin{document}
\title{\textit{Ab initio} four-band Wannier tight-binding model for generic
twisted graphene systems}
\author{Jin Cao}
\affiliation{Key laboratory of advanced optoelectronic quantum architecture and
measurement (MOE), School of Physics, Beijing Institute of Technology,
Beijing 100081, China }
\author{Maoyuan Wang}
\affiliation{Key laboratory of advanced optoelectronic quantum architecture and
measurement (MOE), School of Physics, Beijing Institute of Technology,
Beijing 100081, China }
\affiliation{International Center for Quantum Materials, School of Physics, Peking
University, Beijing 100871, China}

\author{Cheng-Cheng Liu}
\email{ccliu@bit.edu.cn}

\affiliation{Key laboratory of advanced optoelectronic quantum architecture and
measurement (MOE), School of Physics, Beijing Institute of Technology,
Beijing 100081, China }
\author{Yugui Yao}
\email{ygyao@bit.edu.cn}

\affiliation{Key laboratory of advanced optoelectronic quantum architecture and
measurement (MOE), School of Physics, Beijing Institute of Technology,
Beijing 100081, China }
\date{\today}
\begin{abstract}
The newly realized twisted graphene systems such as twisted bilayer
graphene (TBG), twisted double bilayer graphene (TDBG), and twisted
trilayer graphene (TTG) have attracted widespread theoretical attention.
Therefore, a simple and accurate model of the systems is of vital
importance for the further study. Here, we construct the symmetry-adapted
localized Wannier functions and the corresponding \textit{ab initio}
minimal two-valley four-band effective tight-binding models for generic
twisted graphene systems with small twist angle. Such two-valley model
evades the Wannier obstruction caused by the fragile topology in one-valley
model. The real space valley operator is introduced to explicitly
describe the valley $U_{v}\left(1\right)$ symmetry. Each symmetry-adapted
Wannier orbital shows a peculiar three-peak form with its maximum
at AA spots and its center at AB or BA spots. An extended Hubbard
model is also given and the related parameters are presented explicitly.
We provide an approach to systematically build the Wannier tight-binding
model for generic twisted graphene systems. Our model provides a firm
basis for further study of the many-body effects in these systems.
\end{abstract}
\maketitle

\paragraph{\textcolor{blue}{Introduction.}\textemdash{}}

The recent discovery of correlated insulating states and possibly
unconventional superconductivity in magic-angle twisted bilayer graphene
(TBG)\citep{nature2018-Y.Cao-expTBG-Correlated,nature2018-Y.Cao-expTBG-Unconv.sc}
has triggered broad interest in TBG systems\citep{N2019-AliYazdani-expTBG,nature2019-AbhayN.Pasupathy-expTBG,nature2019-Eva.Y.Andrei-expTBG,np2019-YoungjoonChoi-expTBG,prl2019-Tomarken-expTBG,npg2019-Ashvin-weekcouple-SO4-corrTBG,prb2019-ShaffiqueAdam-GaugePhonon-corrTBG,prl2018-FanYang-weekcouple-CSD-corrTBG,prl2019-EnricoRossi-superfluid-corrTBG,prx2018-FuLiang-weekcouple-SC-DW-corrTBG,prl2019-XiaoYanXu-ValenceBondOrders-corrTBG,acsnano2020-HeLin-expTBG,nature2018-Y.Cao-expTBG-Correlated,nature2018-Y.Cao-expTBG-Unconv.sc,prb2017-HeLin-expTBG,np2020-saito-expTDBG,arxiv2020-liu-expTBG,prb2018-LiujunZou-TBG-fragileTB,prb2018-Fuliang-wannierTB,prx2018-FuLiang-wannierTB,prx2018-Kangjian-wannierTB,prb2019-PoHoiChun-tenband-fragileTB,prRes2019-Carr-fragileTB,prx2019-Bohm-JungYang-bandTBG,prl2019-SongZD-alltopo-TBG-fragileTB,prb2019-LiuJianpeng-landaulevel-bandTBG,prl2019-Ashvin-bandTBG,prb2018-HuaimingGuo-corrTBG,prb2019-BitanRoy-corrTBG,prb2019-CenkeXu-corrTBG,prl2018-WuFengcheng-corrTBG,prl2019-Kohn-Luttinger-superconductivity-corrTBG,prx2018-HoiChunPo-corrTBG,sciBulletin2019-TianxingMa-quantumMC-corrTBG,arxiv2019-liushang-corrTBG,arxiv2020-LuChen-corrTBG,prx2020-Ashvin-corrTBG,arxiv2020-JianpengLiu-corrTBG,prb2020-FuchunZhang-corrTBG,prl2020-MingXie-corrTBG}.
Immediately after the magic-angle TBG, twisted double bilayer graphene
(TDBG)\citep{nature2020-G.Y.Zhang-expTDBG,nature2020-Liu-expTDBG,nature2020-Y.Cao-expTDBG,prb2019-Jeil.Jung-expTDBG,prl2019-EmanuelTutuc-expTDBG,prb2019-Choi-bandTDBG,prb2019-koshino-bandTDBG,nc2019-lee-theory-corrTDBG,prb2020-Wu-corrTDBG}
and twisted multilayer graphene (TMG)\citep{prb2018-HeLin-TwoTwistAngles-TTG,arxiv2020-ShiYanmeng-OneTwistAngle-expTTG,prb2018-AdrianaVela-SKTB-TMG,prx2019-Daixi-TMG-M+N,arxiv2019-lixiao-TwoTwistAngles-TTG,prl2020-ZhuZiyan-TwoTwistAngles-TTG}
as well as other twisted two-dimensional materials\citep{np2019-FengWang-3+1-ABC-TLG-hBN-1,prl2019-JeilJung-theory-3+1-TLG-hBN,nanoletter2013-kang-LinWangWang-DFTtwist-TMDCs,prb2019-Koshino-SKTB-TMDCs,prl2018-MacDonald-TMD-Hubbard,nanoletter2019-AngelRubio-exp-TBBN}
have been fabricated and investigated, forming a new research field\textemdash twistronics.
Generic TBG systems are usually described by effective continuum model\citep{pnas2011-Bistritzer-MacDonald-ContinuumModel,prb2011-E.J.Mele-ContinuumModel,prl2007-Neto-ContinuumModel},
tight-binding (TB) model\citep{nanoletter2012-Chou.M.Y.-SKTB-TBG-LandauLevel,prb2010-Barticevic-SKTB-TBG,prb2010-Pankratov-SKTB-TBG,prb2018-AdrianaVela-SKTB-TMG,prb2018-Sboychakov-SKTB-TBG,prb2018-XianqingLin-SKTB-TBG,prb2019-Koshino-SKTB-TMDCs}
and density functional theory (DFT)\citep{jms2014-Correa-DFTtwist-graphene,nanoletter2013-kang-LinWangWang-DFTtwist-TMDCs,nanoletter2019-Lxian-DFTtwist-hBN,prb2014-Uchida-DFTtwist-graphene,prb2017-PengKang-DFTtwist-BlackPhosphorus,prb2019-Lucignano-DFTtwist-graphene}.
However, these models need lots of basis to model the single-particle
band structure. Even for the continuum model, hundreds of basis are
needed, and the other two need tens of thousands and even more, which
has seriously hindered the study of the novel many-body quantum states
in twisted graphene systems. 

For TBG with small twist angle, two kinds of effective Wannier TB
models with several orbitals were proposed\citep{prx2018-FuLiang-wannierTB,prb2018-Fuliang-wannierTB,prx2018-Kangjian-wannierTB,prb2018-LiujunZou-TBG-fragileTB,prb2019-PoHoiChun-tenband-fragileTB,prRes2019-Carr-fragileTB}.
One is building the effective Wannier TB model for one valley, which
we refer to 1V-TB model. Due to the so-called fragile topology in
the low-energy bands, more deliberated-selected extra trivial orbitals
should be added to form a wannierizable group\citep{prb2018-LiujunZou-TBG-fragileTB,prb2019-PoHoiChun-tenband-fragileTB,prl2019-SongZD-alltopo-TBG-fragileTB,prx2019-Bohm-JungYang-bandTBG}.
Obviously, this method cannot be directly promoted to generic twisted
graphene systems. The other is considering two valleys together named
2V-TB model\citep{prb2018-Fuliang-wannierTB,prx2018-FuLiang-wannierTB,prx2018-Kangjian-wannierTB}.
The low-energy flat bands for 2V-TB model in TBG represent a trivial
band topology\citep{prb2019-PoHoiChun-tenband-fragileTB,prl2019-SongZD-alltopo-TBG-fragileTB}.
The 1V-TB model separates two valleys, which explicitly preserves
the $U_{v}\left(1\right)$ valley symmetry, while such symmetry might
be lost in 2V-TB model. 

Here, we attempt to provide a numerical method to systematically construct
\textit{ab initio} minimal four-band Wannier TB models for arbitrary
stacked graphene systems with a small twist angle. First, from \textit{ab
initio} calculations, the band structures of untwisted multilayer
graphene subsystems are obtained. Our \textit{ab initio} calculations
naturally take into account the trigonal warping around the graphene
valleys induced by interlayer coupling\citep{prl2006-FewLayerGraphene-wraping}.
It was often ignored in previous studies and should be included since
the energy scales for the trigonal warping and the flat bands after
twist are comparable. We numerically explicitly demonstrate the chiral
decomposition rule for TMG, and identify the low energy bands and
the distribution in real space. Given this knowledge, the Wannier
functions (WFs) are constructed by combining the microscopic $p_{z}$
orbitals with an envelope function. The TB model for TBG(1+1), TDBG(2+2),
twisted trilayer graphene (TTG)(1+2) and TMG(4+4) are constructed
as examples. Based on the WFs for these systems with small twist angle,
we explicitly present the real space Hamiltonian and the valley operator,
which together encode the whole low-energy physics and the symmetry,
especially the $U_{v}\left(1\right)$ valley symmetry. Finally, the
electron-electron interactions between the localized WFs is discussed
and the extended Hubbard model is given. 

\paragraph{\textcolor{blue}{Chiral decomposition of few-layer and twisted multilayer
graphene.}\textemdash{}}

\begin{figure}
\begin{centering}
\includegraphics[scale=0.45]{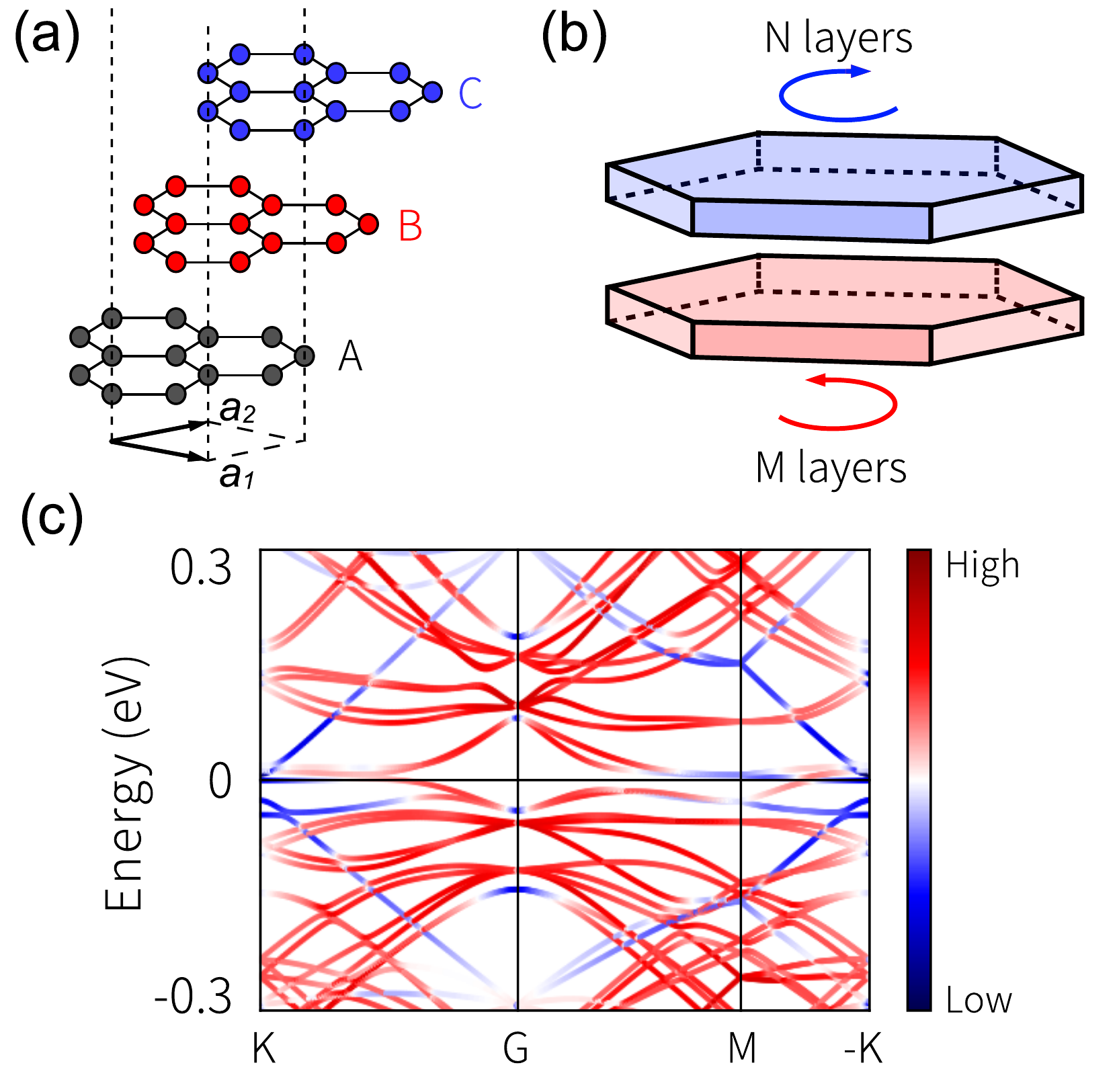}
\par\end{centering}
\centering{}\caption{\label{FIG.1}(a) Stacked nonequivalent layers in few-layer graphene
labeled as A, B and C respectively. (b) Generic twisted multilayer
graphene system with M and N layers arbitrary stacked graphene on
the bottom and top respectively denoted as $\mathrm{TMG_{M}^{N}}$.
(c) Chiral decomposition for $\mathrm{TMG_{M}^{N}}$, where M=\{ABCA\}
and N=\{ABCA\}\{C\}. The band structures are projected onto two nearest
active chiral subsets, i.e., the \{ABCA\} chiral subset on the bottom
and the \{ABCA\} chiral subset on the top.}
\end{figure}

Stacking single-layer graphene (SLG) along $\mathbf{\hat{z}}$ direction
forms a few-layer graphene (FLG) system. The energetically favorable
stacking order is generated by intralayer translations along $\left(\mathbf{a}_{1}+\mathbf{a}_{2}\right)/3$
with an additional interlayer $d_{0}\hat{\boldsymbol{z}}$ resulting
in three non-equivalent layers labeled by A, B and C respectively\citep{RMP2009-graphene-review},
as shown in Fig.\ref{FIG.1}(a). Here $d_{0}$ represents the layer
distance of FLG. One stacking order is referenced as chirally stacking
order if all of the intralayer translations are the same. Clearly
a general stacking sequence can be decomposed into several subsets
of chirally stacking order, the so-called chiral decomposition in
FLG\citep{prb2008-chiral-decomp}: The low energy states of the N-layer
stacked graphene can be well described by direct sum of $N_{\mathrm{D}}$
subspaces, $H_{N}^{\mathrm{eff}}\thickapprox H_{J_{1}}\oplus\cdots\oplus H_{J_{N_{\mathrm{D}}}}$,
where each of $H_{J_{i}}$ is a pseudospin doublet with $k^{J_{i}}$
leading order dispersion induced from the \textit{i}th chiral subset
in N-layer stacked graphene with the sum rule $\sum_{i=1}^{N_{\mathrm{D}}}J_{i}=N$.
Furthermore, these low energy states are localized at the boundary
of the chirally stacking subsets. For example, a chirally stacking
$J_{i}$-layer subset has $J_{i}-1$ pairs of dimmer and two unpaired
sites left. The two unpaired sites contribute two zero modes, which
are responsible for the low energy subspace, and the rest parts are
pushed into high energy by strong direct interlayer coupling. The
previous studies usually considered the ideal case, where are only
the nearest interlayer hopping parameters taken into account, and
ignored the trigonal warping and particle-hole (PH) asymmetry in realistic
case. We would like to point out that the effects should be included
because of the comparable energy scale with that of the low-energy
flat bands. In this work, the electronic structure for FLG are obtained
from \textit{ab initio} calculations, where the trigonal warping and
PH asymmetry are automatically included (see Appendix.\ref{sec:SUP.comput.method}
for details). The \textit{ab initio} results show a well preserved
chiral decomposition rule as presented in Appendix.\ref{sec:SUP.FLG}.

Considering a general TMG system, as presented in Fig.\ref{FIG.1}(b),
it has N-layer graphene on the top and M-layer graphene on the bottom
with small twist angles $\pm\theta/2$ respectively. In the absence
of twisted interlayer coupling, this system is described by several
groups of pseudospin doublets with $k^{J_{i}}$ leading order dispersion.
With the twisted interlayer coupling turned on, in fact two active
chiral subsets are responsible for the flat bands: the bottom chiral
subset of the upper N layers and the top chiral subset of the lower
M layers. From the continuum model, we numerically calculated the
orbital characters for TMG system. We choose $\mathrm{TMG_{\{ABCA\}}^{\{ABCAC\}}}$
with a relative twist angle $2.646^{\circ}$as an example. It can
be decomposed into three chiral subsets \{ABCA\}, \{ABCA\} and \{C\}
in quartic and linear dispersion. As presented in Fig.\ref{FIG.1}(c),
the former two active twisted chiral subsets strongly renormalize
into the flat bands, and the left one preserves well the linear dispersion.
In brief, the low-energy states in TMG can be well described by two
decoupled parts: the renormalized flat bands and the left pseudospin
doublet. 

\paragraph{\textcolor{blue}{Four-band Wannier tight-binding model for generic
TMG.}\textemdash{}}

\begin{figure}
\begin{centering}
\includegraphics[scale=0.65]{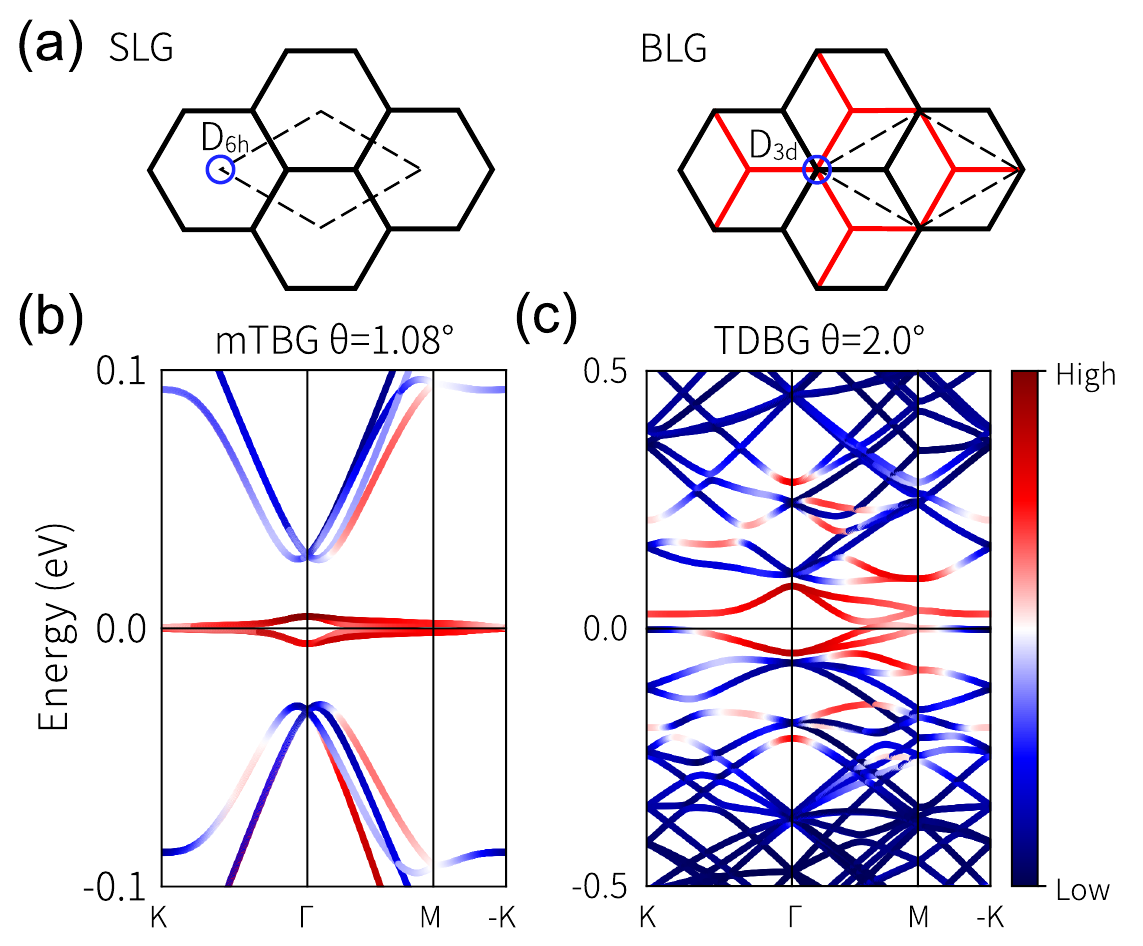}
\par\end{centering}
\caption{\label{FIG.2}(a) Maximum site symmetry for SLG and BLG as circled
in blue. The black and red solid lines stand for layers. (b) and (c)
Band projections onto the initial WFs for TBG and TDBG ($\mathrm{TMG_{\{AB\}}^{\{AB\}}}$).
The flat bands can be well described by the initial WFs by means of
the large overlap. And it is guaranteed since these WFs notably overlap
with the zero modes, which are responsible for the low energy states.}
\end{figure}

We start by discussing the symmetry of TMG. The space groups of all
FLG are symmorphic, i.e., apart from the lattice translations, all
of the symmetric operations leave one site fixed {[}see Fig.\ref{FIG.2}(a){]}.
When twisted, the symmetry group depends on the twist center. To get
the maximal symmetric structure in real space, one should take the
twist center at that of the maximal site symmetry. The symmetry of
the band structure is not sensitive to this atomic level selection
since the moir\'{e} pattern is much larger than the atomic length
scale. The nonzero twist angle removes the inversion symmetry. Based
on this knowledge, the maximal symmetry groups for the twisted graphene
systems are given in Table.\ref{TABLE.1}. The $\mathcal{C}_{3z}$
symmetry is in general preserved. 

In the absence of twisted interlayer coupling, the low energy Bloch
states in TMG can be viewed as folding the band structure of FLG,
which are induced from the microscopic $p_{z}$ orbitals of FLG. Also,
these low energy Bloch states come from the graphene valleys thus
taking a high frequency factor $e^{i\mathbf{K}_{\xi}^{\mathrm{FLG}}\cdot\mathbf{r}}$,
where $\mathbf{K}_{\xi}^{\mathrm{FLG}}$ is the valley of FLG. It
suggests that the WFs can be explicitly written in the form

\begin{eqnarray}
\left|g_{n}\right\rangle  & = & \frac{1}{2}\sum_{\xi;\tau,d,\mathbf{R}}e^{i\mathbf{K}_{\xi}^{\mathrm{FLG}}\cdot\mathbf{r}}f_{n}^{\left(\xi;\tau,d\right)}\left(\mathbf{r}\right)\left|\tau,d,\mathbf{R}+\mathbf{d}\right\rangle .
\end{eqnarray}
Here $\left|\tau,d,\mathbf{R}+\mathbf{d}\right\rangle $ is the microscopic
$p_{z}$ orbitals of FLG and the sum runs over valley index $\xi$,
sublattice $\tau$, layer index $d$ and graphene lattice $\mathbf{R}$.
$n$ is the index of the WFs, and $\tau=\tau_{\alpha},\tau_{\beta}$
denotes the position of the sublattice. $f_{n}^{\left(\xi;\tau,d\right)}$
is the smooth envelope function in moir\'{e} length scale. Since we
consider a 2V-TB model, the entire system preserves the time reversal
symmetry. It is possible to choose a group of real-valued WFs with
constraint $f_{n}^{\left(\xi_{+};\tau,d\right)}=f_{n}^{\left(\xi_{-};\tau,d\right)}\equiv f_{n}^{\left(\tau,d\right)}$.
It should be pointed out that different from the 1V-TB model, in which
the valley degrees of freedom are promoted as orbitals, in the 2V-TB
model, the valley degrees of freedom are denoted as orbital components.
Also, our choice of WFs equally mixes two valleys. This mixture cannot
be removed due to the non-trivial topology of the flat bands from
a single valley. Despite of this choice, it is possible to preserve
well the valley $U_{v}\left(1\right)$ symmetry as we will discuss
later. 

\begin{table}
\begin{centering}
\caption{\label{TABLE.1}Symmetry of the twisted multilayer graphenes.}
\par\end{centering}
\begin{ruledtabular}
\begin{centering}
\begin{tabular}{ccc}
Stacking order & Point Group & Generators\tabularnewline
\hline 
TBG & $D_{6}$ & $\mathcal{C}_{6z},\mathcal{C}_{2x}$\tabularnewline
$\mathrm{TMG_{\{AB\}}^{\{AB\}}}$ (TDBG) & $D_{3}$ & $\mathcal{C}_{3z},\mathcal{C}_{2x}$\tabularnewline
$\mathrm{TMG_{\{AB\}}^{\{BA\}}}$ (TDBG) & $D_{3}$ & $\mathcal{C}_{3z},\mathcal{C}_{2x}$\tabularnewline
generic $\mathrm{TMG_{N}^{M}}$ & $C_{3}$ & $\mathcal{C}_{3z}$\tabularnewline
\end{tabular}
\par\end{centering}
\end{ruledtabular}
\end{table}

For generic TMG, we choose the envelope function for the initial WFs
as $f_{1}^{\left(\tau_{\alpha},-1\right)}\left(\mathbf{r}\right)=G\left(\mathbf{r}-\mathbf{r}_{1}^{\mathrm{hex}}\right)$,
$f_{2}^{\left(\tau_{\beta},-1\right)}\left(\mathbf{r}\right)=G\left(\mathbf{r}-\mathbf{r}_{2}^{\mathrm{hex}}\right)$,
$f_{3}^{\left(\tau_{\beta},1\right)}\left(\mathbf{r}\right)=-G\left(\mathbf{r}-\mathbf{r}_{1}^{\mathrm{hex}}\right)$
and $f_{4}^{\left(\tau_{\alpha},1\right)}\left(\mathbf{r}\right)=-G\left(\mathbf{r}-\mathbf{r}_{2}^{\mathrm{hex}}\right)$.
Here $G\left(\mathbf{r}-\mathbf{r}_{i}^{\mathrm{hex}}\right)$ is
the Gaussian function localized at the hexagonal site with moir\'{e}
scale spreading. The initial choice is based on the following considerations.
Firstly, it reflects the realistic orbitals character since these
chosen WFs have notable overlap with the relevant flat bands, as explicitly
shown in Fig.\ref{FIG.2}. Secondly, the initial choice respects the
corresponding symmetry: (i) Each $g_{n}$ itself has $\mathcal{C}_{3z}$
symmetry due to the site symmetry at hexagonal site. The $\mathcal{C}_{3z}$
symmetry is preserved for all TMG systems. (ii) In each graphene layer,
$\mathcal{C}_{2z}$ exchanges the sublattice degrees of freedom, thus
exchanges $g_{1}$, $g_{3}$ with $g_{2}$, $g_{4}$ respectively.
The $\mathcal{C}_{2z}$ symmetry is only preserved in TBG. (iii) $\mathcal{C}_{2x}$
exchanges $g_{1}$, $g_{2}$ with $g_{3}$, $g_{4}$ respectively.
The $\mathcal{C}_{2x}$ symmetry is preserved in TBG and TDBG. 

\begin{figure}
\begin{centering}
\includegraphics[scale=0.9]{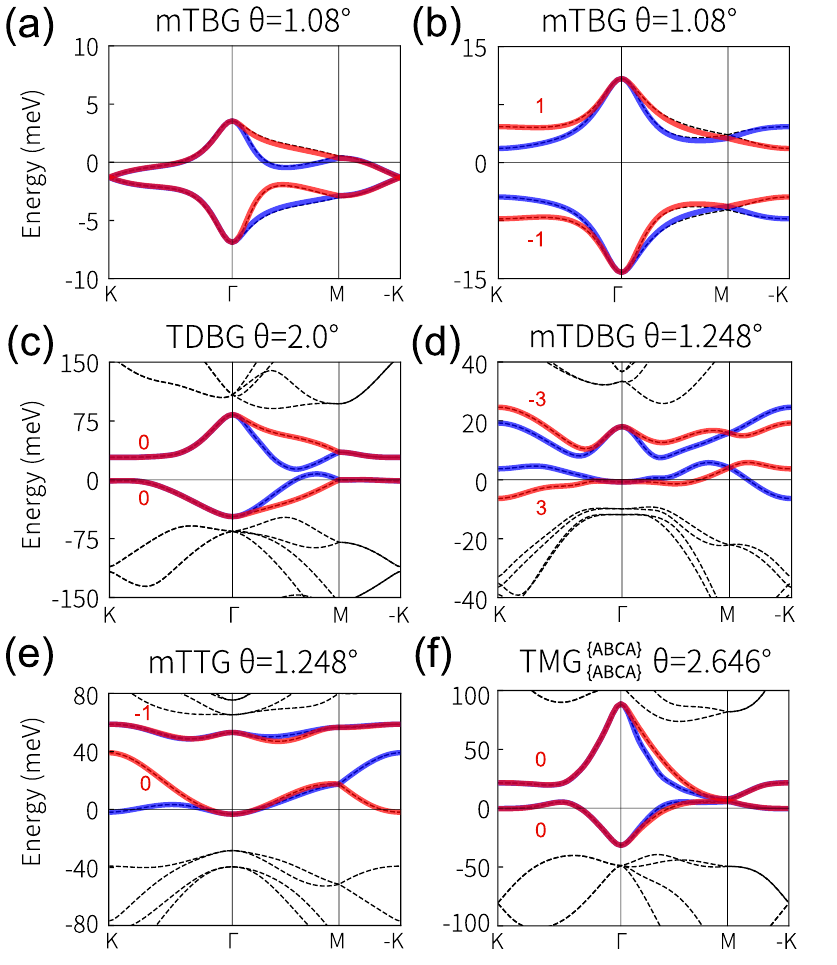}
\par\end{centering}
\caption{\label{FIG.3}Band structures interpolated from Wannier tight-binding
model (in red and blue lines) for (a) TBG, (b) TBG on h-BN substrate,
(c) TDBG, (d) magic angle TDBG, (e) magic angle TTG and (f) $\mathrm{TMG_{\{ABCA\}}^{\{ABCA\}}}$
respectively, with the effects of atomic relaxations taken in to account.
As a comparison, the band structures by combining the effective continuum
model and \textit{ab initio} calculation are given in black dash lines.
Our Wannier tight-binding band structures are obtained by simultaneously
diagonalizing the Hamiltonian and the valley operator, and thus can
be labeled with a well-defined valley eigenvalues $\pm1$ $\left(\mathbf{K},\mathbf{K}^{\prime}\right)$
highlighted in red and blue lines.}
\end{figure}

\begin{figure*}
\begin{centering}
\includegraphics[scale=0.8]{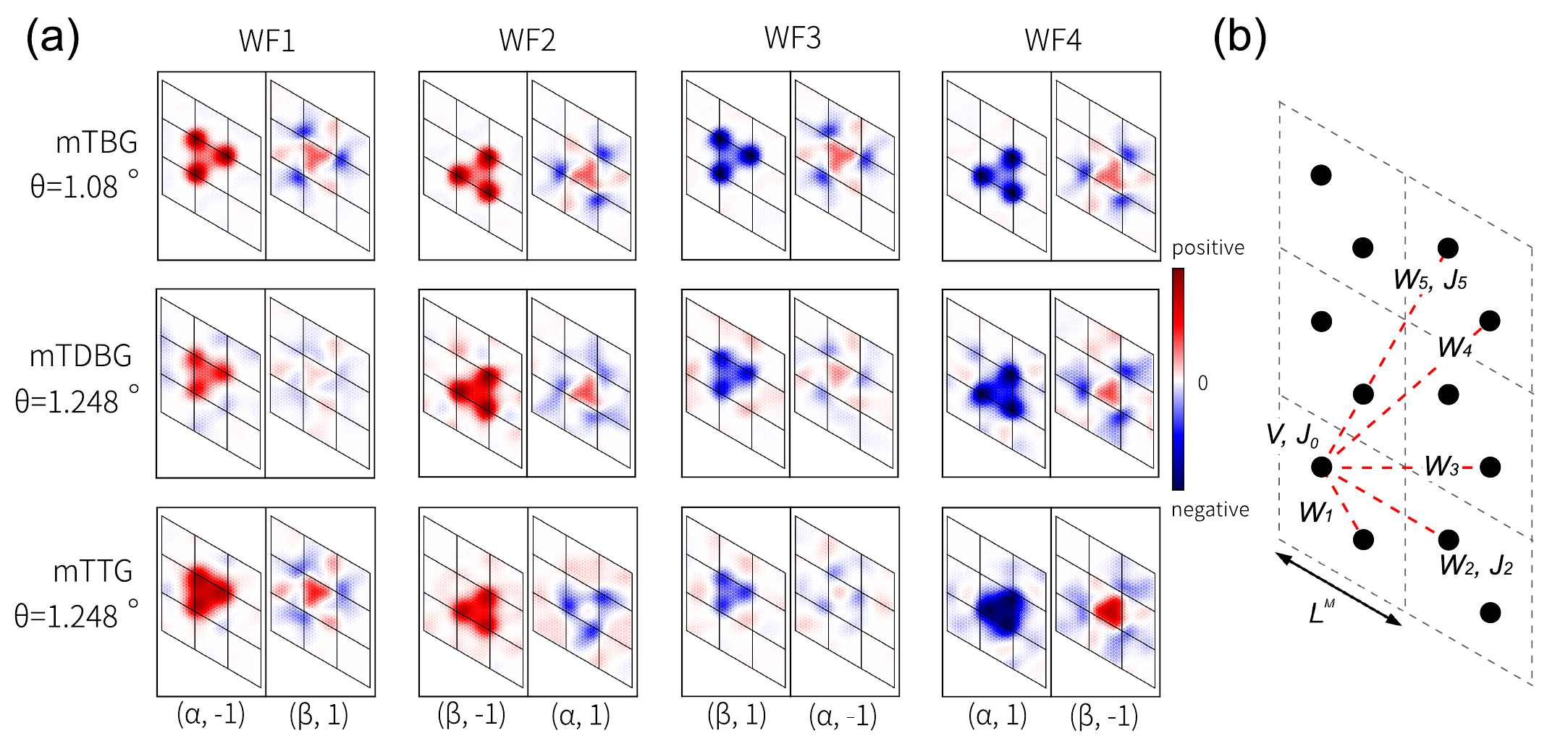}
\par\end{centering}
\caption{\label{FIG.4}(a) Plot of the envelope part $f_{n}^{\left(\xi;\tau,d\right)}\left(\mathbf{r}\right)$
of $C_{3z}$-symmetry-adapted localized WFs for mTBG with magic angle
$\theta=1.08^{\circ}$, mTDBG with magic angle $\theta=1.248^{\circ}$
and mTTG with magic angle $\theta=1.248^{\circ}$ respectively. Only
$\mathbf{K}$ valley case is presented, the another one is the same.
The $3\times3$ moir\'{e} cell is shown in black lines. (b) Nonzero
Coulomb interactions between these WFs. The extended Hubbard interactions
are taken into account due to the extended features of the WFs. For
exchange interactions, the nonzero terms presented here are $J_{0}$,
$J_{2}$ and $J_{5}$. The other terms, i.e., $J_{1}$, $J_{3}$ and
$J_{4}$ are interactions between the two WFs, which are located at
different layers or sublattices resulting in zero exchange interactions.
Notice that WFs have three peaks at the moir\'{e} triangular lattice
sites (a) but are centered at the dual honeycomb lattice sites (b).}
\end{figure*}

Given these envelope functions, one can construct the WFs and the
related TB model following the methodology developed by D. Vanderbilt
et al. \citep{WANNIER90_composite_1997,WANNIER90_entangled_2001}
(see Appendix.\ref{sec:SUP.wannierTB} for details). The Wannier TB
models and the WFs for $\mathrm{TMG_{N}^{M}}$ are automatically produced
by our home-made code. With the TB model of the few-layer graphene
from \textit{ab initio} calculations (see Appendix.\ref{sec:SUP.comput.method}),
one can readily obtain the four-band Wannier TB model for generic
TMG with small twist angle. Here we build TB models for several prototypical
TMG systems for examples, i.e., mTBG with twist angle $1.08^{\circ}$,
mTBG with magic angle $1.08^{\circ}$ and h-BN substrate $\Delta_{\textrm{BN}}^{d=-1}=30\,\textrm{meV}$,
TDBG with twist angle $2.0^{\circ}$, mTDBG with magic angle $1.248^{\circ}$
and displacements field tuned on ($U_{\textrm{D}}=10\,\textrm{meV}$),
mTTG ($\mathrm{TMG_{\{A\}}^{\{AB\}}}$) with magic angle $1.248^{\circ}$
($U_{\textrm{D}}=80\,\textrm{meV}$, $\Delta_{\textrm{BN}}^{d=-1}=80\,\textrm{meV}$
and $\Delta_{\textrm{BN}}^{d=1}=\Delta_{\textrm{BN}}^{d=2}=-20\,\textrm{meV}$),
and $\mathrm{TMG_{\{ABCA\}}^{\{ABCA\}}}$ with twist angle $2.646^{\circ}$.
They band structures are shown in Fig.\ref{FIG.3}, which fit well
with the effective continuum model. The $\mathcal{C}_{3z}$ symmetry
is enforced in building symmetry-adapted WFs since it is preserved
in all TMG system, and also reflected in the hopping parameters between
WFs, as plotted in Figs.\ref{FIG.S3}-\ref{FIG.S5}. The WFs show
a peculiar three-peak form, as shown in Fig.\ref{FIG.4}(a). Beside
the one component of each Wannier orbital we set up initially, there
emerges another component located at same hexagonal site but in different
layer and sublattice. It should be pointed out this is different from
the previous studies\citep{prx2018-FuLiang-wannierTB,prx2018-Kangjian-wannierTB},
where each component is nonzero for all of the WFs. Our Wannier TB
model well describes the Berry curvature distributions and Chern numbers
of generic TMG, e.g., TBG, TDBG and so on, with zero valley Chern
number, i.e., the sum of Chern number of the two flat bands for single
valley is zero. The Berry curvature distributions and Chern numbers
for both valleys are given in Appendix.\ref{sec:SUP.chern}. Although
the topological description of TMG with nonzero valley Chern number
is beyond our current Wannier TB model, the model fits the four flat
bands very well for generic TMG with or without nonzero valley Chern
number. As an example of TMG with nonzero valley Chern number, mTTG
has zero Chern number for the valence band and nonzero one for the
conduction band, which suggests one may include higher energy bands
rather than the four flat bands to characterize the topological aspect.

An important feature for twisted graphene systems with small twist
angle is the well preserved valley $U_{v}\left(1\right)$ symmetry
due to the negligible intervalley coupling. The $U_{v}\left(1\right)$
symmetry is explicitly present in continuum model. However, such symmetry
does not seem to exist in our two-valley Wannier TB model. In fact,
our method indeed manifests well the $U_{v}\left(1\right)$ symmetry
since the initial Bloch sum is unitarily transformed from the original
flat bands obtained by the continuum model {[}see Eq.(\ref{eq:SUP.EQ.unitary_transision}){]}.
To reveal the valley degrees of freedom in our 2V-TB model, we define
the valley operator for the continuum model, and the valley operator
for our Wannier TB model can be interpolated from the exactly same
procedure as that of Hamiltonian as illustrated in Appendix.\ref{sec:SUP.wannierTB}.
The eigenvalues of interpolated valley operator are stabilized at
$\pm1$ as shown in Fig.\ref{FIG.S.ValleyEignValues}. By simultaneously
diagonalizing the Hamiltonian and the valley operator, the bands and
eigenstates in our Wannier TB model can be labeled with valley eigenvalues,
and thus the $U_{v}\left(1\right)$ symmetry is preserved. The bands
with different valley eigenvalues $\pm1$ $\left(\mathbf{K},\mathbf{K}^{\prime}\right)$
are colored in blue and red in Fig.\ref{FIG.3}. 

\paragraph{\textcolor{blue}{The extended Hubbard model for generic TMG.}\textemdash{}}

\begin{table}
\begin{centering}
\caption{\label{TABLE.2}Coulomb interaction between WFs in unit of $\frac{e^{2}}{4\pi\epsilon L^{\mathrm{M}}}$,
which are integrated in a $5\times5$ moir\'{e} lattice. Taking $\epsilon_{r}=5$
for a BN substrate, the unit for mTBG ($\theta=1.08^{\circ}$) is
$22\,\mathrm{meV}$ and are $26\,\mathrm{meV}$ for mTDBG ($\theta=1.248^{\circ}$)
and mTTG ($\theta=1.248^{\circ}$) . The $\mu$ and $\bar{\mu}$ represent
for different orbitals.}
\par\end{centering}
\begin{ruledtabular}
\begin{centering}
\begin{tabular}{cccc}
 & mTBG($1.08^{\circ}$) & mTDBG($1.248^{\circ}$) & mTTG($1.248^{\circ}$)\tabularnewline
\hline 
$U$ & 2.210 & 1.991 & 2.330\tabularnewline
$V$ & 2.208 & 2.011 & 2.150\tabularnewline
$W_{1}$ & 1.742 & 1.636 & 1.689\tabularnewline
$W_{2}$ & 1.238 & 1.151 & 1.149\tabularnewline
$W_{3}$ & 1.160 & 1.053 & 1.029\tabularnewline
$W_{4}$ & 0.700 & 0.761 & 0.717\tabularnewline
$W_{5}$ & 0.623 & 0.653 & 0.615\tabularnewline
$J_{0}^{\mu\bar{\mu}}$ & 0.000 & 0.016 & 0.253\tabularnewline
$J_{2}^{\mu\mu}$ & 0.192 & 0.117 & 0.070\tabularnewline
$J_{2}^{\mu\bar{\mu}}$ & 0.044 & 0.036 & 0.049\tabularnewline
$J_{5}^{\mu\mu}$ & 0.002 & 0.017 & 0.002\tabularnewline
$J_{5}^{\mu\bar{\mu}}$ & 0.003 & 0.006 & 0.006\tabularnewline
\end{tabular}
\par\end{centering}
\end{ruledtabular}
\end{table}

Finally, we discuss the electron-electron interactions between the
constructed localized WFs. Because each Wannier orbital has two nonzero
components out of the total four components and shows the peculiar
three-peak form, the interaction Hamiltonian takes the following form 

{\small{}
\begin{eqnarray}
H_{int} & = & U\sum_{i\mu}n_{i\mu\uparrow}n_{i\mu\downarrow}+V\sum_{i\mu>\nu}n_{i\mu}n_{i\nu}\nonumber \\
 &  & +\frac{1}{2}\sum_{\alpha}W_{\alpha}\sum_{\left\langle ij\right\rangle _{\alpha},\mu\nu}n_{i\mu}n_{j\nu}\nonumber \\
 &  & +\sum_{i\mu>\nu,\sigma\sigma^{\prime}}J_{0}a_{i\mu\sigma}^{\dagger}a_{i\nu\sigma}a_{i\nu\sigma^{\prime}}^{\dagger}a_{i\mu\sigma^{\prime}}\nonumber \\
 &  & +\sum_{i\mu\neq\nu}J_{0}a_{i\mu\uparrow}^{\dagger}a_{i\nu\uparrow}a_{i\mu\downarrow}^{\dagger}a_{i\nu\downarrow}\nonumber \\
 &  & +\frac{1}{2}\sum_{\left\langle ij\right\rangle _{2}\mu\nu,\sigma\sigma^{\prime}}J_{2}^{\mu\nu}a_{i\mu\sigma}^{\dagger}a_{j\nu\sigma}a_{j\nu\sigma^{\prime}}^{\dagger}a_{i\mu\sigma^{\prime}}\nonumber \\
 &  & +\sum_{\left\langle ij\right\rangle _{2}\mu\nu}J_{2}^{\mu\nu}a_{i\mu\uparrow}^{\dagger}a_{j\nu\uparrow}a_{i\mu\downarrow}^{\dagger}a_{j\nu\downarrow}
\end{eqnarray}
}where $a_{i\mu\sigma}^{\dagger}$ creates a WFs at $i$ site with
spin index $\sigma$. $\mu=g_{1},g_{3}$ or $g_{2},g_{4}$ depending
on the location of the Wannier center. The numerical integration values
of the parameters in the interaction Hamiltonian for mTBG, mTDBG and
mTTG are summarized in Table.\ref{TABLE.2}. The extended Hubbard
interactions are represented by the $W_{\alpha}$ ($\alpha\!=\!1\!\sim\!5$)
For the first three terms, the quantitively difference is mainly determined
by the distance between the localized WFs. The two WFs, which are
in the same shape but located in different layers or sublattice, have
little effect on the direct interactions. For Hund's exchange and
pair-hopping terms, the nonzero terms are shown in Fig.\ref{FIG.4}(b).
It is nonzero only for the WFs located at the same hexagonal site
up to the lattice translations. The reason is that, for instance the
nonzero components of $g_{1}$ and $g_{2}$ are located at different
layers or sublattices resulting in a zero exchange interaction. 

\paragraph{\textcolor{blue}{Conclusion and discussion.}\textemdash{}}

We present an approach to construct $\mathcal{C}_{3z}$-symmetry-adapted
localized WFs and the corresponding \textit{ab initio} minimal four-band
effective tight-binding models for generic twisted graphene systems
with small twist angle, such as experimentally realized TBG, TDBG,
and TTG, as well as other $\mathrm{TMG_{M}^{N}}$ systems experimentally
to be realized. Each symmetry-adapted Wannier orbital shows a peculiar
three-peak form with two nonzero components out of the total four
components. An extended Hubbard model is also obtained and the related
parameters are calculated explicitly.

Our starting point is the \textit{ab initio} band structures of untwisted
multilayer graphene subsystems, which naturally includes the trigonal
warping effect, which was often ignored in previous study and should
be taken into account since the energy scales for the trigonal warping
and the flat band after twist are comparable. Then we use the continuum
model to address the twisted cases with the lattice relaxation effect
taken into account. For a single-valley model, constructing two-band
Wannier model will have an obstruction due to the fragile topology
in TBG systems, but for a two-valley four-band model, it is wannierable.
Moreover, the valley $U_{v}\left(1\right)$ symmetry can still be
retrieved in our two-valley four-band model. We can use the eigenvalue
of the valley operator $\pm1$ $\left(\mathbf{K},\mathbf{K}^{\prime}\right)$
to mark the energy band by constructing the valley operator explicitly,
and then diagonalizing it with the Hamiltonian simultaneously. The
effect of external electric displacement field and sublattice symmetry
breaking can be readily incorporated in our model. Our \textit{ab
initio} minimal four-band effective Wannier tight-binding models together
with the extended interactions is of importance for a wide range of
applications in the efficient study of the many-body effects in the
TMG systems.
\begin{acknowledgments}
C.-C. L thanks F. Yang, L.-D. Zhang for earlier related collaborations
on TBG, and J. Liu for fruitful discussions. This work was supported
by the NSF of China (Grants Nos. 11922401, 11734003, 11774028, 11574029),
the National Key R\&D Program of China (Grant No. 2016YFA0300600),
the Strategic Priority Research Program of Chinese Academy of Sciences
(Grant No. XDB30000000).
\end{acknowledgments}

\bibliographystyle{apsrev4-2}
\bibliography{ref,ref_method}

\appendix
\begin{center}
\onecolumngrid
\setcounter{figure}{0}
\par\end{center}

\section{Calculation methods \label{sec:SUP.comput.method}}

In this work, the electronic structures for $\mathrm{TMG_{M}^{N}}$
are obtained from the effective continuum method\citep{pnas2011-Bistritzer-MacDonald-ContinuumModel}.
The band structures for the untwisted N-layer and M-layer FLG are
adapted from the \textit{ab initio} results (see detail in Appendix.\ref{sec:SUP.BMmodel}).
The \textit{ab initio} calculations for FLG were performed in the
VASP package\citep{VASP} and the electron-ion interaction was described
using the projector augmented wave (PAW) method\citep{PAW}. The exchange-correlation
part was described with the generalized gradient approximation (GGA)\citep{GGA}
in the scheme of Perdew-Burke-Ernzerhof (PBE) functional\citep{PBE}.
The plane-wave cutoff energy was set to be 400 eV. The Brillouin zone
(BZ) was sampled by a $\Gamma$ centered Monkhorst-Pack grid ($12\times12\times1$)\citep{MPSample}.
The lattice constant for all FLG is set at 2.46 $\mathrm{\mathring{A}}$.
The Wannier tight-binding model for FLG was constructed by the WANNIER90
code\citep{WANNIER90_CPC}. 

\section{Orbital character of few-layer graphene \label{sec:SUP.FLG}}

\begin{figure}
\renewcommand{\thefigure}{S.\arabic{figure}}
\begin{centering}
\includegraphics{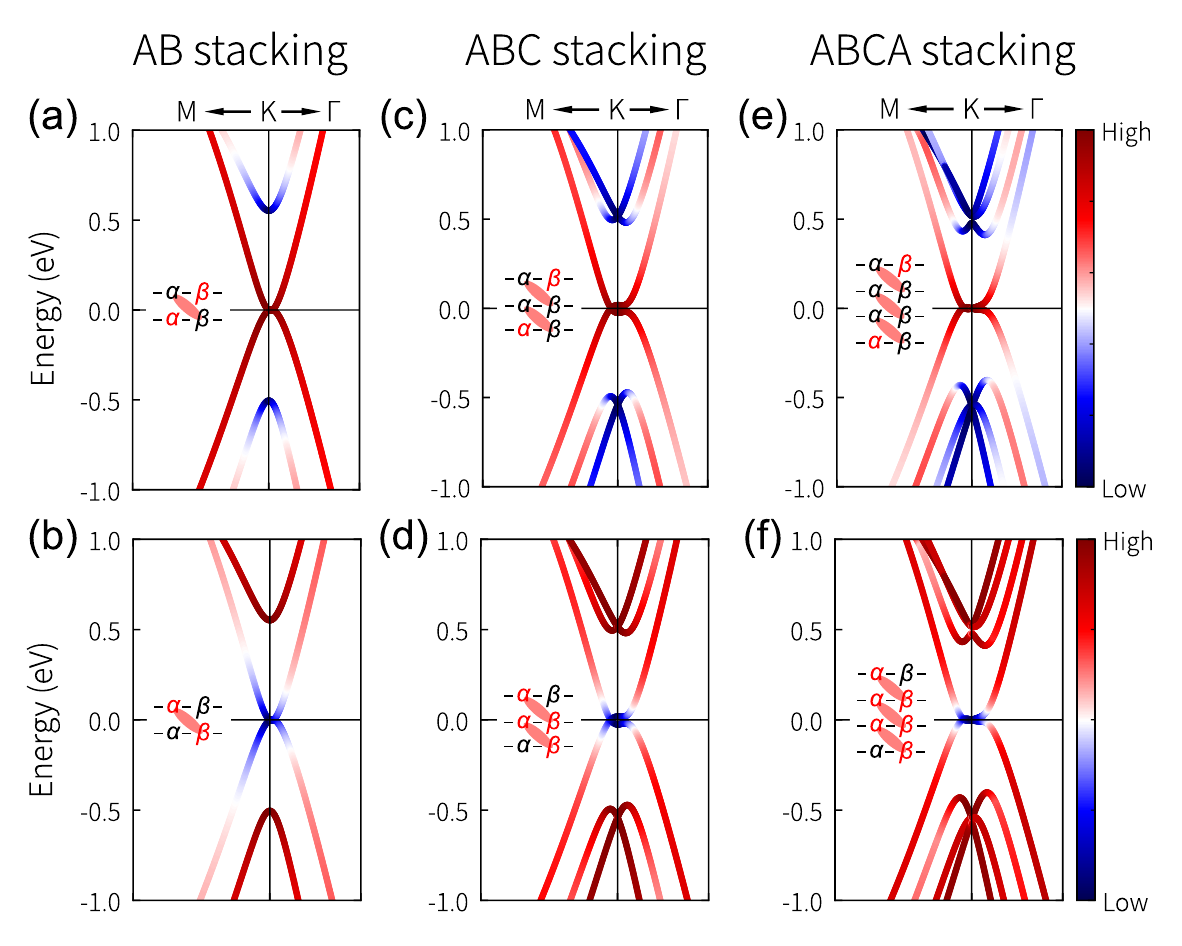}
\par\end{centering}
\caption{\label{FIG.S1}Projected band structures near $\mathbf{K}$ valley
for chirally stacked few-layer graphene with stacking order AB, ABC
and ABCA respectively. The selected $p_{z}$ orbitals are colored
in red as shown inside each subfigure. (a),(c) and (e) show the projection
on the top and bottom layer, which mainly dominated the low energy
states. (b), (d) and (f) show the projection on the rest part, which
is almost distributed at high energy states.}
\end{figure}

In this section, we present the orbitals character for FLG with either
chirally and multi-chirally stacking order from \textit{ab initio}
calculations. The band projections around $\mathbf{K}$ valley for
chirally stacked FLG are presented in Fig.\ref{FIG.S1}. The low energy
states are well described by the pseudospin doublet with approximately
quadratic, cubic and quartic dispersion for AB, ABC and ABCA stacking
order respectively. The Fermi surface wrapping effects are automatically
taken into account. By projecting onto the $p_{z}$ orbitals at zero
mode sites {[}highlight in red in Fig.\ref{FIG.S1}(a), (c) and (e){]},
we found the low energy states, which dominate the physics in the
absence of twist angle, are mainly distributed in these zero mode
sites. The contributions of the rest $p_{z}$ orbitals are far away
from the Fermi energy. 

\begin{figure}
\renewcommand{\thefigure}{S.\arabic{figure}}
\begin{centering}
\includegraphics{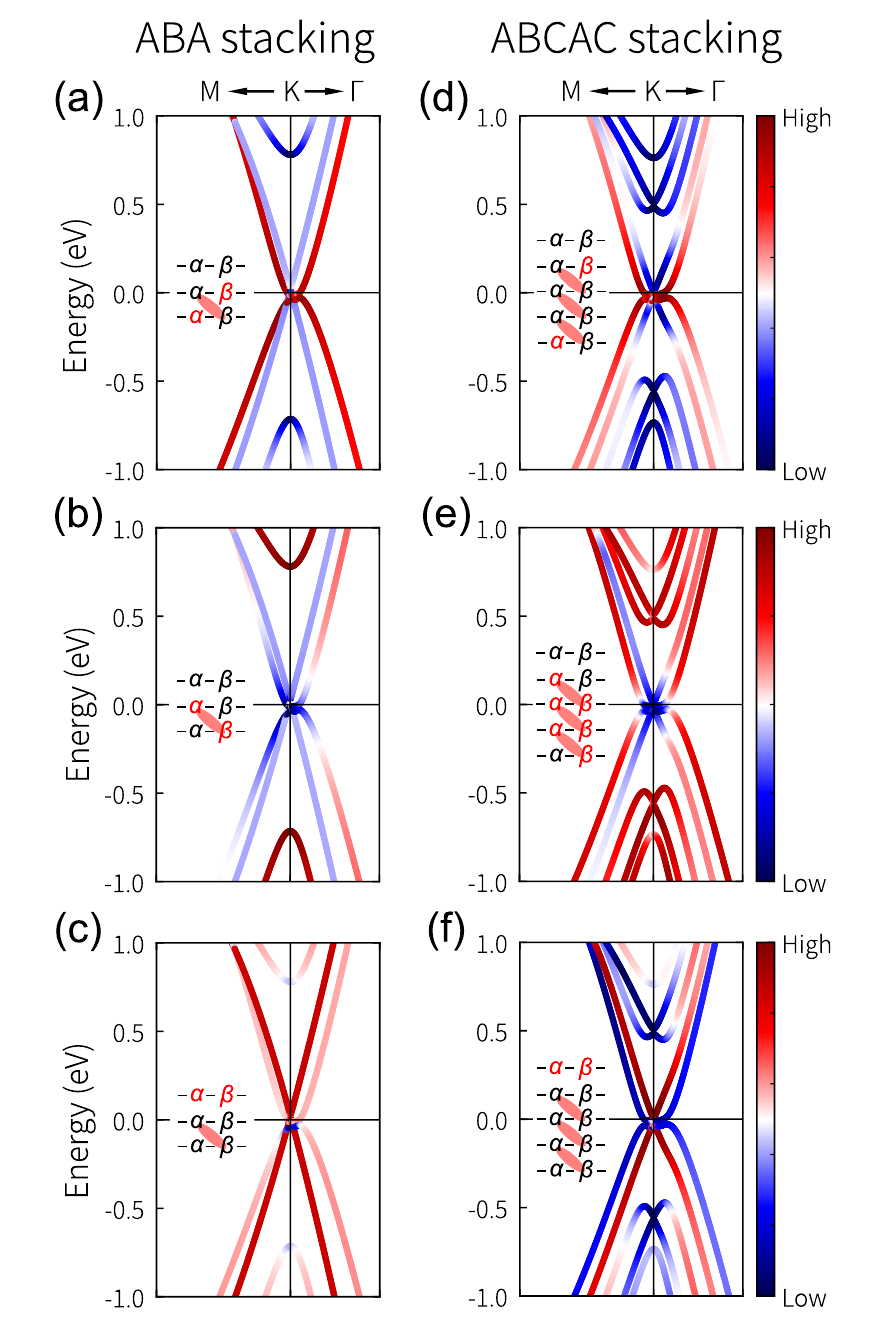}
\par\end{centering}
\caption{\label{FIG.S2}Projected band structures near $\mathbf{K}$ valley
for multi-chirally stacked few-layer graphene with stacking order
ABA and ABCAC respectively. The selected $p_{z}$ orbitals are colored
in red as illustrated inside each subfigure.}
\end{figure}

For multi-chirally stacked FLG, we numerically calculated ABA and
ABCAC stacked FLG, which can be decomposed into \{AB\}\{A\} and \{ABCA\}\{C\}
respectively. The band structures near $\mathbf{K}$ valley are illustrated
in Fig.\ref{FIG.S2}. The low energy states for ABA stacked FLG are
two pseudospin doublets in linear and quadratic dispersion respectively,
and the rest bands are pushed into high energy indicating a well preserved
chiral decomposition rule despite a generic hopping parameters are
included in our \textit{ab initio} results. Again, by projecting onto
the $p_{z}$ orbitals at zero mode sites {[}highlight in red in Fig.\ref{FIG.S2}(a)
and (c){]}, we found they mainly contribute to the two doublets respectively.
The rest $p_{z}$ orbitals contribute to the high energy bands. Similar
results are obtained for ABCAC stacked FLG as shown in Fig.\ref{FIG.S2}. 

In summary, the chiral decomposition rules are well preserved in the
\textit{ab initio} results. Moreover, the low energy pseudospin doublets
in FLG, which play an important role in the presence of twist angle,
are mainly contributed from the $p_{z}$ orbitals localized in the
zero mode sites. The Fermi surface wrapping effect are automatically
taken into account. 

\section{The effective continuum model for generic twisted graphene systems
\label{sec:SUP.BMmodel}}

In this section, we illustrate the combination of \textit{ab initio}
Wannier TB model for FLG and the effective continuum method\citep{pnas2011-Bistritzer-MacDonald-ContinuumModel}
used in this work to obtain the single particle electronic structures
for generic twisted multilayer graphene system (TMG). The geometric
structure is illustrated in Fig.\ref{FIG.1} in the main text. There
are N layers graphene on the top and M layers graphene on the bottom
with small twist angles $\pm\theta/2$ respectively. The emerged moir\'{e}
pattern is labeled as $\mathbf{L}^{\mathrm{M}}=l_{1}\mathbf{L}_{1}^{\mathrm{M}}+l_{2}\mathbf{L}_{2}^{\mathrm{M}}$.
The Bloch states are constructed from the microscopic $p_{z}$ orbitals
in each layer of graphene $\left|\tau,d,\mathbf{L}^{\mathrm{M}}+\mathbf{R}+\mathbf{d}\right\rangle $.
Here $\tau=\tau_{\alpha},\tau_{\beta}$ represents graphene sublattice
degree of freedom and $d$ is the layer index measured from the bottom
to the top (for TDBG, $d=-2,-1,1,2$ from the bottom to the top layer).
$\mathbf{R}$ represents the graphene unit cell (in each moir\'{e}
pattern). $\mathbf{d}=dd_{0}\mathbf{\hat{z}}$ represent the layer
stacking distance where $d_{0}$ is the distance between two graphene
layers. The Bloch sum functions read as 
\begin{eqnarray}
\left|\tilde{\psi}_{\mathbf{k}_{\xi}^{d}+\mathbf{G}}^{\left(\xi;\tau_{\alpha},d\right)}\right\rangle  & = & \sum_{\mathbf{L}^{\mathrm{M}},\mathbf{R}}e^{i\left(\mathbf{k}_{\xi}^{d}+\mathbf{G}\right)\cdot\mathcal{D}\left[\mathrm{sign}\left(d\right)\frac{\theta}{2}\right]\left(\mathbf{L}^{\mathrm{M}}+\mathbf{R}+\mathbf{\tau}_{\alpha}\right)}\left|\mathbf{\tau}_{\alpha},d,\mathbf{L}^{\mathrm{M}}+\mathbf{R}+\mathbf{d}\right\rangle .\label{eq:SUP.EQ.blochsum}
\end{eqnarray}
The summations run over all graphene unit cell in Born-von Karman
supercell. Here $\xi\equiv\xi_{\pm}=\pm1$ represents different graphene
valleys. $\mathbf{k}_{\xi}^{d}=\mathbf{k}+\mathbf{K}_{\xi}^{\mathrm{FLG}}-\mathbf{K}_{\xi}^{d}$
is measured from the $\Gamma$ point in the graphene Brillouin zone,
and $\mathbf{k}$ is measured from the $\Gamma$ point in the TMG
Brillouin zone. $\mathbf{K}_{\xi}^{\textrm{FLG}}=\frac{\xi}{3}\left(2\mathbf{G}_{1}^{\textrm{FLG}}+\mathbf{G}_{2}^{\textrm{FLG}}\right)$
is the graphene valley, $\mathbf{K}_{\xi}^{d>0}=\frac{\xi}{3}\left(\mathbf{G}_{1}^{\textrm{M}}-\mathbf{G}_{2}^{\textrm{M}}\right)$
and $\mathbf{K}_{\xi}^{d<0}=-\frac{\xi}{3}\left(\mathbf{G}_{1}^{\textrm{M}}+2\mathbf{G}_{2}^{\textrm{M}}\right)$.
$\mathbf{G}=n_{1}\mathbf{G}_{1}^{\textrm{M}}+n_{2}\mathbf{G}_{2}^{\textrm{M}}$
is the reciprocal lattice vector for TMG, $\mathcal{D}\left[\mathrm{sign}\left(d\right)\frac{\theta}{2}\right]$
indicates a twisted angle $\theta/2$ for the top part and $-\theta/2$
for the bottom part. The low energy states can be expended as the
Bloch sum functions near the two valleys

\begin{eqnarray}
\left|\psi_{n\mathbf{k}}\right\rangle  & = & \sum_{\xi;\tau,d}\sum_{\mathbf{G}}C_{n\mathbf{k}}^{\left(\xi;\tau,d\right)}\left(\mathbf{G}\right)\left|\tilde{\psi}_{\mathbf{k}_{\xi}^{d}+\mathbf{G}}^{\left(\xi;\tau,d\right)}\right\rangle \nonumber \\
 & = & \sum_{X\mathbf{G}}C_{n\mathbf{k}}^{X}\left(\mathbf{G}\right)\left|\tilde{\psi}_{\mathbf{k}_{\xi}^{d}+\mathbf{G}}^{X}\right\rangle ,
\end{eqnarray}
where we have rewritten $X=\left(\xi;\tau,d\right)$ for simplicity.
$C_{n\mathbf{k}}^{X}\left(\mathbf{G}\right)$ can be obtained by diagonalizing
the effective continuum model

\begin{eqnarray}
\hat{H}\left(k\right) & = & \hat{H}_{0}\left(k\right)+\hat{H}_{\textrm{T}}+\hat{H}_{\textrm{BN}}+\hat{H}_{\textrm{D}},\label{eq:SUP.EQ.BM.Hk}
\end{eqnarray}
where $\hat{H}_{0}$ describes the few-layer graphene in the top and
bottom part, and $\hat{H}_{\textrm{T}}$ is the effective twisted
interlayer coupling. The $\mathcal{C}_{2z}$ symmetry can be removed
by considering the effect of h-BN substrate $\hat{H}_{\textrm{BN}}$.
And $\hat{H}_{\textrm{D}}$ describes the displacement field to separate
the flat bands. In the Bloch sum basis Eq.(\ref{eq:SUP.EQ.blochsum}),
the Hamiltonian matrix elements read as

\begin{eqnarray}
\left\langle \tilde{\psi}_{\mathbf{k}_{\xi}^{d}+\mathbf{G}}^{\left(\xi;\tau_{\alpha},d\right)}|\hat{H}_{0}|\tilde{\psi}_{\mathbf{k}_{\xi^{\prime}}^{d^{\prime}}+\mathbf{G}^{\prime}}^{\left(\xi^{\prime};\tau_{\beta},d^{\prime}\right)}\right\rangle  & = & \delta_{\xi\xi^{\prime}}\delta_{\mathrm{sign}\left(d\right),\mathrm{sign}\left(d^{\prime}\right)}\delta_{\mathbf{G}\mathbf{G}^{\prime}}\nonumber \\
 &  & \times\sum_{\mathbf{R}}e^{i\left(\mathbf{k}_{\xi}^{d}+\mathbf{G}\right)\cdot\mathbf{R}}\left\langle \mathbf{\tau}_{\alpha},d,\mathbf{0}+\mathbf{d}|\hat{H}_{0}|\mathbf{\tau}_{\beta},d^{\prime},\mathbf{R}+\mathbf{d}^{\prime}\right\rangle ,
\end{eqnarray}
which are obtained by Fourier transforming the Wannier tight-binding
model of FLG. The matrix elements for $\hat{H}_{\textrm{BN}}$ are

\begin{eqnarray}
\left\langle \tilde{\psi}_{\mathbf{k}_{\xi}^{d}+\mathbf{G}}^{\left(\xi;\tau_{\alpha},d\right)}|\hat{H}_{\textrm{BN}}|\tilde{\psi}_{\mathbf{k}_{\xi^{\prime}}^{d^{\prime}}+\mathbf{G}^{\prime}}^{\left(\xi^{\prime};\tau_{\beta},d^{\prime}\right)}\right\rangle  & = & \delta_{\xi\xi^{\prime}}\delta_{dd^{\prime}}\delta_{\mathbf{G}\mathbf{G}^{\prime}}\sigma_{\alpha\beta}^{z}\Delta_{\textrm{BN}}^{d}.
\end{eqnarray}
The matrix elements for $\hat{H}_{\textrm{D}}$ are 

\begin{eqnarray}
\left\langle \tilde{\psi}_{\mathbf{k}_{\xi}^{d}+\mathbf{G}}^{\left(\xi;\tau_{\alpha},d\right)}|\hat{H}_{\textrm{D}}|\tilde{\psi}_{\mathbf{k}_{\xi^{\prime}}^{d^{\prime}}+\mathbf{G}^{\prime}}^{\left(\xi^{\prime};\tau_{\beta},d^{\prime}\right)}\right\rangle  & = & \delta_{\xi\xi^{\prime}}\delta_{dd^{\prime}}\delta_{\mathbf{G}\mathbf{G}^{\prime}}\times\sigma_{\alpha\beta}^{0}U_{\textrm{D}}^{d},
\end{eqnarray}
where we define $U_{\textrm{D}}\equiv U_{\textrm{D}}^{\max\left(d\right)}-U_{\textrm{D}}^{\min\left(d\right)}$.
The twisted interlayer coupling read as

\begin{eqnarray}
\left\langle \tilde{\psi}_{\mathbf{k}_{\xi}^{d}+\mathbf{G}}^{\left(\xi;\tau_{\alpha},d\right)}|\hat{H}_{\textrm{T}}|\tilde{\psi}_{\mathbf{k}_{\xi^{\prime}}^{d^{\prime}}+\mathbf{G}^{\prime}}^{\left(\xi^{\prime};\tau_{\beta},d^{\prime}\right)}\right\rangle  & = & \delta_{\xi\xi^{\prime}}\delta_{d=\pm1,d^{\prime}=\mp1}\nonumber \\
 &  & \times\left[T_{1}\delta_{\mathbf{G},\mathbf{G}^{\prime}}+T_{2}\delta_{\mathbf{G},\mathbf{G}^{\prime}+\xi\mathbf{G}_{1}^{\textrm{M}}}+T_{3}\delta_{\mathbf{G},\mathbf{G}^{\prime}+\xi\left(\mathbf{G}_{1}^{\textrm{M}}+\mathbf{G}_{2}^{\textrm{M}}\right)}\right],
\end{eqnarray}

\begin{eqnarray}
T_{1} & = & \left(\begin{array}{cc}
u & u^{\prime}\\
u^{\prime} & u
\end{array}\right),\quad T_{2}=\left(\begin{array}{cc}
u & u^{\prime}\omega^{-\xi}\\
u^{\prime}\omega^{\xi} & u
\end{array}\right),\quad T_{3}=\left(\begin{array}{cc}
u & u^{\prime}\omega^{\xi}\\
u^{\prime}\omega^{-\xi} & u
\end{array}\right),
\end{eqnarray}
$\omega=e^{2\pi i/3}$. The twisted interlayer coupling parameters
$u=0.0797\,\mathrm{eV}$ and $u^{\prime}=0.0975\,\mathrm{eV}$, which
take the relaxation effect into account\citep{prx2018-FuLiang-wannierTB}.

\section{Building Wannier tight-binding model \label{sec:SUP.wannierTB}}

In this section, we present the details of building the Wannier tight-binding
model for TMG following the methodology built-in WANNIER90\citep{WANNIER90_composite_1997,WANNIER90_entangled_2001}.
As illustrated in the main text, it is possible to choose a group
of WFs $\left\{ \left|g_{n}\right\rangle \right\} $ in moir\'{e}
pattern scales to represent the low energy flat bands
\begin{eqnarray}
\left|g_{n}\right\rangle  & = & \frac{1}{2}\sum_{\xi}\sum_{\tau,d,\mathbf{L}^{\mathrm{M}},\mathbf{R}}e^{i\mathbf{K}_{\xi}^{\mathrm{FLG}}\cdot\mathbf{r}}f_{n}^{\left(\xi;\tau,d\right)}\left(\mathbf{r}\right)\left|\tau,d,\mathbf{L}^{\mathrm{M}}+\mathbf{R}+\mathbf{d}\right\rangle ,
\end{eqnarray}
where the high frequency part is explicitly presented and $f_{n}^{\left(\xi;\tau,d\right)}$
is the smooth envelope function in moir\'{e} length scale. With the
constraint of real-valued WFs, i.e., $f_{n}^{\left(\xi_{+};\tau,d\right)}=f_{n}^{\left(\xi_{-};\tau,d\right)}\equiv f_{n}^{\left(\tau,d\right)}$,
it follows
\begin{eqnarray}
\left|g_{n}\right\rangle  & = & \sum_{\tau,d,\mathbf{L}^{\mathrm{M}},\mathbf{R}}\cos\left(\mathbf{K}_{\xi}^{\mathrm{FLG}}\cdot\mathbf{r}\right)f_{n}^{\left(\tau,d\right)}\left(\mathbf{r}\right)\left|\tau,d,\mathbf{L}^{\mathrm{M}}+\mathbf{R}+\mathbf{d}\right\rangle ,
\end{eqnarray}
which indicates an equal mixture of two valleys. The initial guess
for the Bloch sum functions are obtained by projecting the initial
WFs $g_{n}$ onto the Bloch states of TMG

\begin{eqnarray}
\left|\tilde{\phi}_{n\mathbf{k}}^{\left(0\right)}\right\rangle  & = & \sum_{m}\left|\psi_{m\mathbf{k}}\right\rangle \left\langle \psi_{m\mathbf{k}}|g_{n}\right\rangle .
\end{eqnarray}
The orbital projection matrix $\left\langle \psi_{m\mathbf{k}}|g_{n}\right\rangle $
can be calculated as

\begin{eqnarray}
A_{mn}\left(\mathbf{k}\right) & = & \left\langle \psi_{m\mathbf{k}}|g_{n}\right\rangle =\sum_{X\mathbf{G}}C_{m\mathbf{k}}^{X}\left(\mathbf{G}\right)^{\ast}\left\langle \tilde{\psi}_{\mathbf{k}_{\xi}^{d}+\mathbf{G}}^{X}|g_{n}\right\rangle ,
\end{eqnarray}

\begin{eqnarray}
\left\langle \tilde{\psi}_{\mathbf{k}_{\xi}^{d}+\mathbf{G}}^{X}|g_{n}\right\rangle  & = & \frac{1}{2\pi}\int d\mathbf{r}\,e^{-i\left(\mathbf{k}_{\xi}^{d}+\mathbf{G}\right)\cdot\mathcal{D}\left[\mathrm{sign}\left(d\right)\frac{\theta}{2}\right]\mathbf{r}}\cos\left(\mathbf{K}_{\xi}^{\mathrm{FLG}}\cdot\mathbf{r}\right)f_{n}^{\left(\tau,d\right)}\left(\mathbf{r}\right).
\end{eqnarray}
The last equation is the inner product of the initial Wannier orbital
and the Bloch sum function. We then preform the singular value decomposition
(SVD) to orthogonalize the initial Bloch sum 

\begin{eqnarray}
\left|\tilde{\phi}_{n\mathbf{k}}^{\left(1\right)}\right\rangle  & = & \sum_{m}\left|\psi_{m\mathbf{k}}\right\rangle \left(A_{\mathbf{k}}S_{\mathbf{k}}^{-1/2}\right)_{mn},\label{eq:SUP.EQ.unitary_transision}
\end{eqnarray}
where

\begin{eqnarray}
A_{\mathbf{k}} & = & U_{\mathbf{k}}\Sigma_{\mathbf{k}}V_{\mathbf{k}}^{\dagger}\\
S_{\mathbf{k}}^{-1/2} & = & V_{\mathbf{k}}\frac{1}{\sqrt{\Sigma_{\mathbf{k}}^{\dagger}\Sigma_{\mathbf{k}}}}V_{\mathbf{k}}^{\dagger}.
\end{eqnarray}
To well describe the subspace of flat bands, we project $\tilde{\phi}_{n\mathbf{k}}^{\left(1\right)}$
onto the subspace spanned by flat bands 

\begin{eqnarray}
\left|\tilde{\phi}_{n\mathbf{k}}^{\left(2\right)}\right\rangle  & = & \mathcal{P}_{\mathbf{k}}^{\left(f.b.\right)}\left|\tilde{\phi}_{n\mathbf{k}}^{\left(1\right)}\right\rangle ,\\
\mathcal{P}_{\mathbf{k}}^{\left(f.b.\right)} & = & \sum_{n}\left|\psi_{n\mathbf{k}}^{\left(f.b.\right)}\right\rangle \left\langle \psi_{n\mathbf{k}}^{\left(f.b.\right)}\right|.
\end{eqnarray}
This procedure was first introduced by D. Vanderbilt et al. to well
describe certain range of the Bloch bands\citep{WANNIER90_entangled_2001}.
Once we get the proper Bloch wave function, the Wannier tight-binding
model can be interpolated by

\begin{eqnarray}
H_{mn}\left(\mathbf{R}\right) & = & \frac{1}{N_{\mathbf{k}}}\sum_{\mathbf{k}}e^{-i\mathbf{k}\cdot\mathbf{R}}\left\langle \tilde{\phi}_{m\mathbf{k}}^{\left(2\right)}|E_{\mathbf{k}}I|\tilde{\phi}_{n\mathbf{k}}^{\left(2\right)}\right\rangle ,\label{eq:SUP.EQ.wannierTB.HR}
\end{eqnarray}
where $E_{\mathbf{k}}$ is eigenvalue obtained from the effective
continuum model. 

\begin{figure*}
\renewcommand{\thefigure}{S.\arabic{figure}}
\begin{centering}
\includegraphics[scale=0.8]{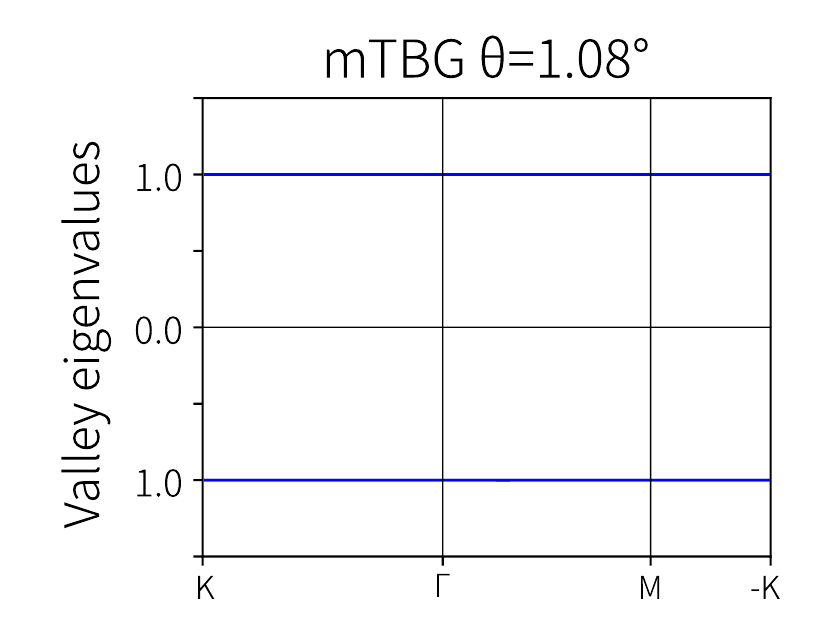}
\par\end{centering}
\caption{\label{FIG.S.ValleyEignValues}Valley eigenvalues interpolated from
real space valley operator of TBG with twist angle $\theta=1.08^{\circ}$.}
\end{figure*}

To clarify the valley degree of freedom in the present framework,
we first define the valley operator in continuum model

\begin{eqnarray}
\mathcal{V}_{\mathbf{k}} & = & \sum_{\xi;\tau,d}\sum_{\mathbf{G}}\xi\left|\tilde{\psi}_{\mathbf{k}_{\xi}^{d}+\mathbf{G}}^{\left(\xi;\tau,d\right)}\right\rangle \left\langle \tilde{\psi}_{\mathbf{k}_{\xi}^{d}+\mathbf{G}}^{\left(\xi;\tau,d\right)}\right|.
\end{eqnarray}
Then, the valley operator for the Wannier TB model can be interpolated
following the exactly same procedure for the Hamiltonian

\begin{eqnarray}
\mathcal{V}_{mn}\left(\mathbf{R}\right) & = & \frac{1}{N_{\mathbf{k}}}\sum_{\mathbf{k}}e^{-i\mathbf{k}\cdot\mathbf{R}}\left\langle \tilde{\phi}_{m\mathbf{k}}^{\left(2\right)}|\mathcal{V}_{\mathbf{k}}|\tilde{\phi}_{n\mathbf{k}}^{\left(2\right)}\right\rangle .\label{eq:SUP.EQ.wannierTB.VR}
\end{eqnarray}
From the real space Hamiltonian $H\left(\mathbf{R}\right)$ and the
valley operator $\mathcal{V}\left(\mathbf{R}\right)$ one can get
the Hamiltonian and the valley operator at certain $\mathbf{k}$ point.
The Hamiltonian can be explicitly classified into two decoupled blocks
labeled with valley eigenvalues $\pm1$ respectively. The valley labeled
Bloch states can be obtained by simultaneously diagonalizing the two
operators. 

\begin{figure*}
\renewcommand{\thefigure}{S.\arabic{figure}}
\begin{centering}
\includegraphics[scale=0.9]{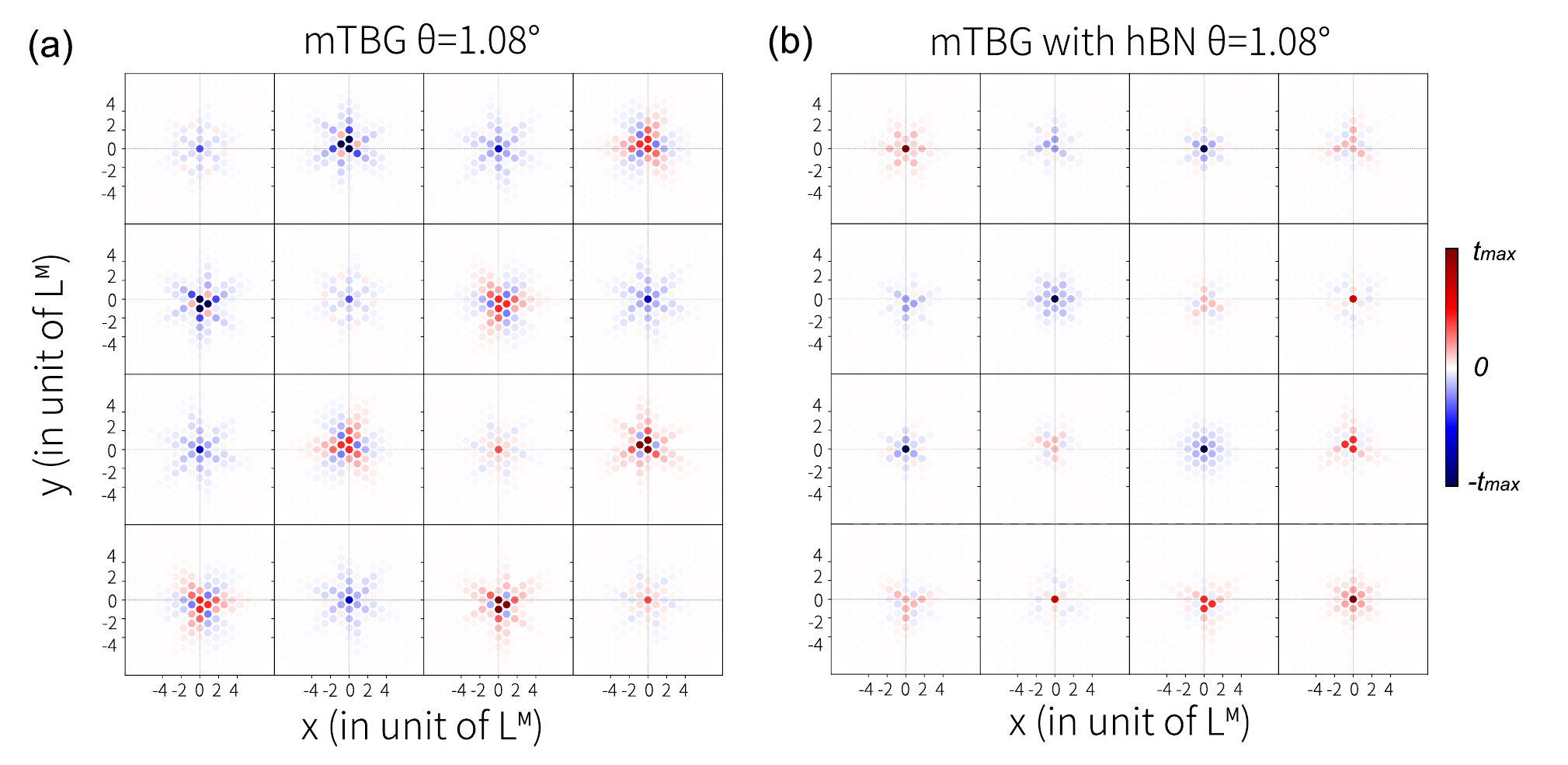}
\par\end{centering}
\caption{\label{FIG.S3}Real space hopping parameters between WFs for (a) the
magic angle TBG, and (b) the magic angle TBG with the effect of h-BN
substrate. The positive values are shown in red whereas the negative
ones are shown in blue. }
\end{figure*}

\begin{figure*}
\renewcommand{\thefigure}{S.\arabic{figure}}
\begin{centering}
\includegraphics[scale=0.9]{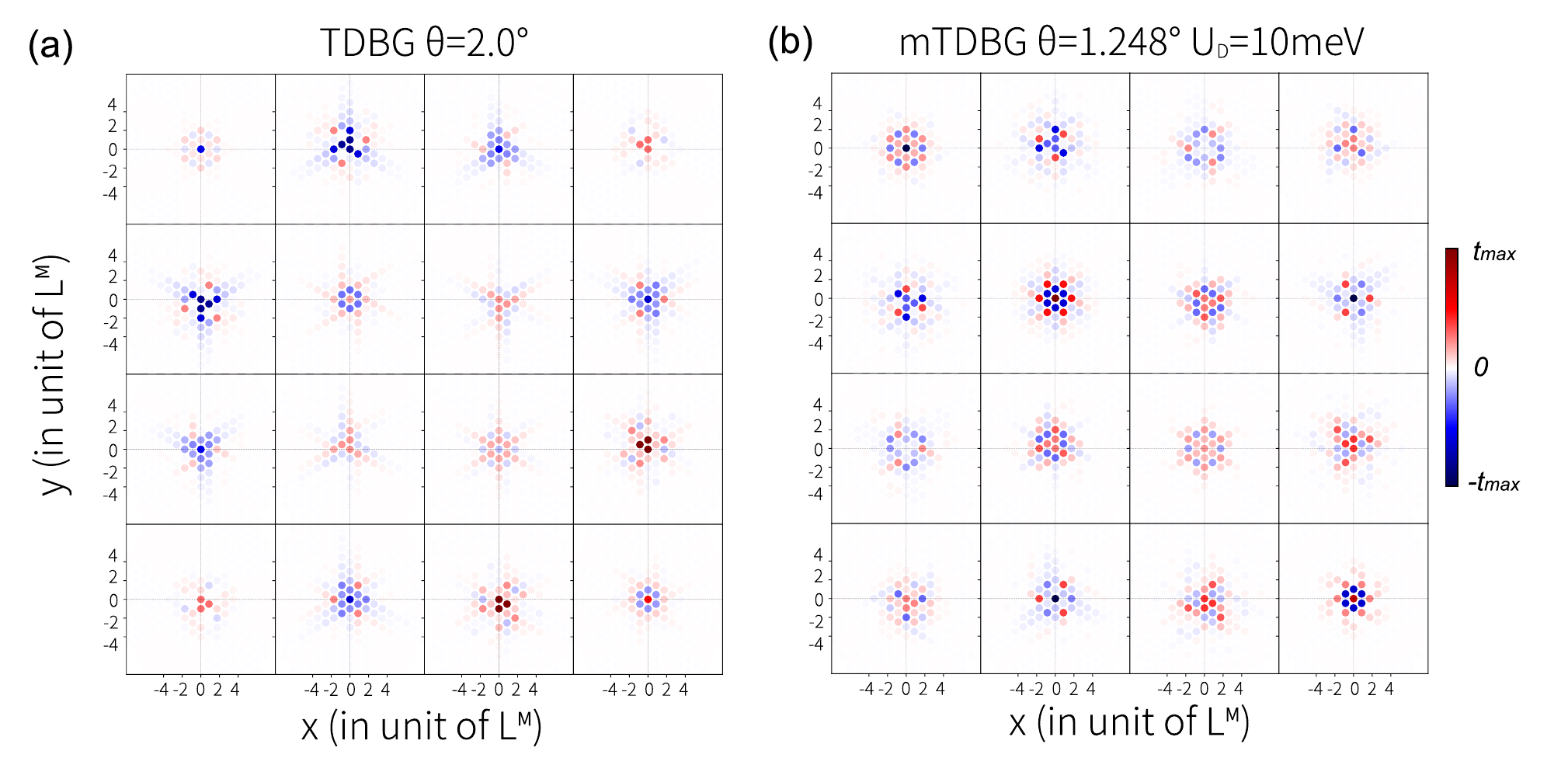}
\par\end{centering}
\caption{\label{FIG.S4}Real space hopping parameters between WFs for (a) the
TDBG, and (b) the magic angle TDBG with the displacement field turned
on ($U_{\textrm{D}}=10\,\textrm{meV}$). The positive values are shown
in red whereas the negative ones are shown in blue. }
\end{figure*}

\begin{figure*}
\renewcommand{\thefigure}{S.\arabic{figure}}
\begin{centering}
\includegraphics[scale=0.9]{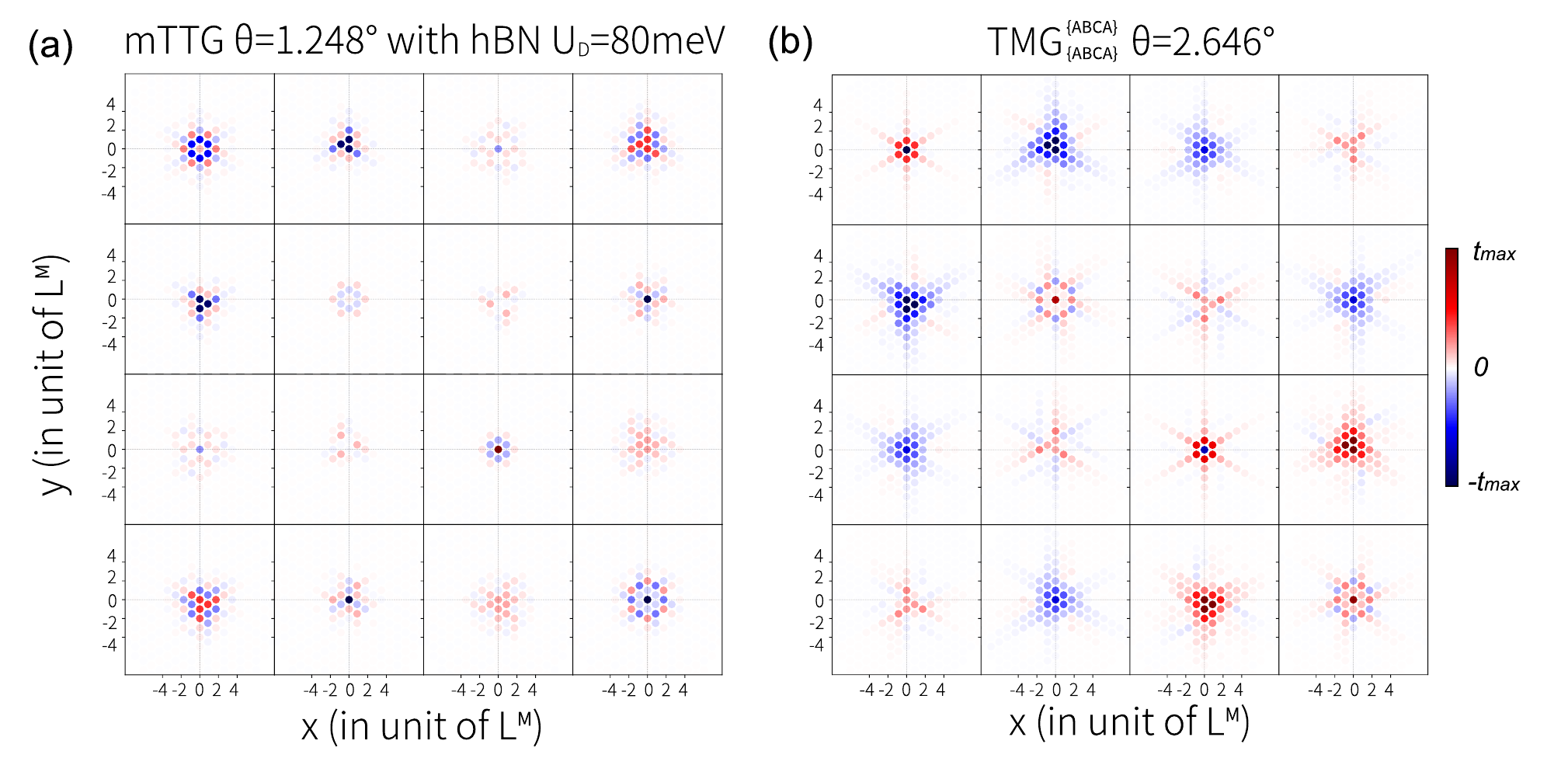}
\par\end{centering}
\caption{\label{FIG.S5}Real space hopping parameters between WFs for (a) the
magic angle TTG ($\mathrm{TMG_{\{A\}}^{\{AB\}}}$) with h-BN substrate
and the displacement field turned on ($U_{\textrm{D}}=80\,\textrm{meV}$,
$\Delta_{\textrm{BN}}^{d=-1}=80\,\textrm{meV}$ and $\Delta_{\textrm{BN}}^{d=1}=\Delta_{\textrm{BN}}^{d=2}=-20\,\textrm{meV}$),
and (b) the $\mathrm{TMG_{\{ABCA\}}^{\{ABCA\}}}$. The positive values
are shown in red whereas the negative ones are shown in blue. }
\end{figure*}

A uniform $18\times18$ mesh for the Brillouin zone were used for
interpolate all of the Wannier tight-binding model in this work. The
hopping parameters are real numbers since the WFs are real-valued
functions. The $\mathcal{C}_{3z}$ symmetry is enforced to build WFs
and so do the hopping parameters. The Hamiltonian matrix elements
are shown in Fig.\ref{FIG.S3}, Fig.\ref{FIG.S4} and Fig.\ref{FIG.S5}.
The resulting Wannier tight-binding models are well documented and
available in GitHub\citep{twistwanTB}. It serves as a start point
for further study of many-body effects in TMG systems. The eigenvalues
for the valley operator of TBG with twist angle $\theta=1.08^{\circ}$
are shown in Fig.\ref{FIG.S.ValleyEignValues}. It is stabilized in
$\pm1$. Similar results are obtained for other presented systems. 

\section{Chern number for continuum model and Wannier TB model \label{sec:SUP.chern}}

\begin{figure*}
\renewcommand{\thefigure}{S.\arabic{figure}}
\begin{centering}
\includegraphics[scale=0.8]{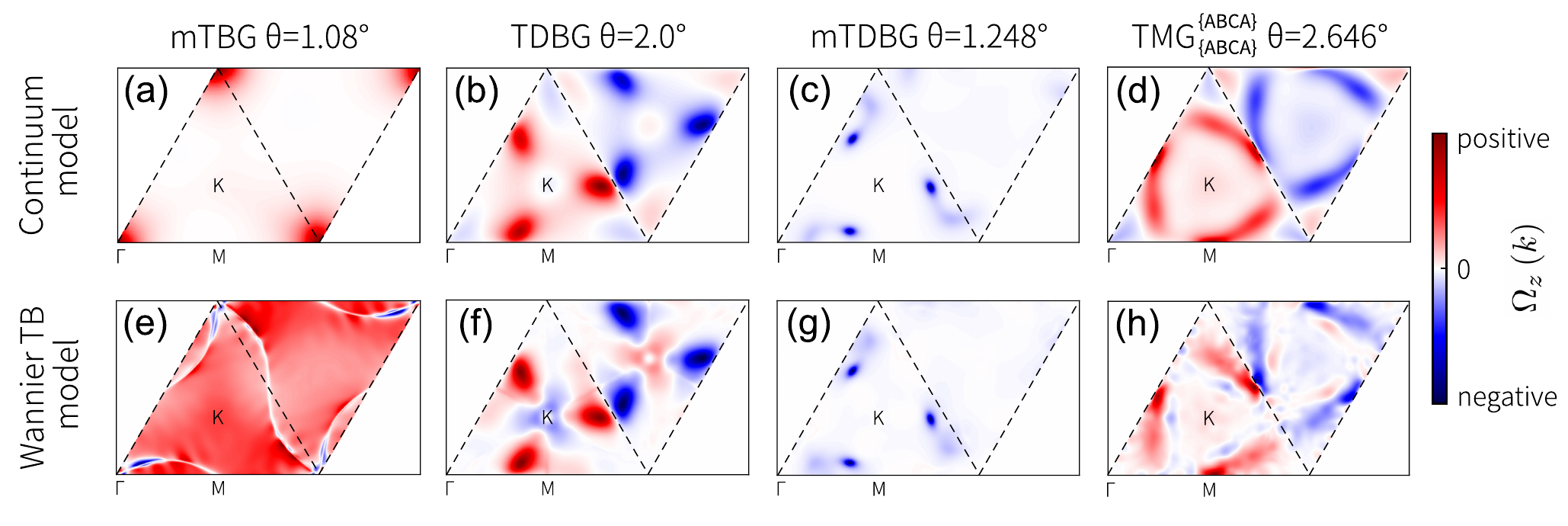}
\par\end{centering}
\caption{\label{FIG.S6.bcdist}Berry curvature distributions for the valence
flat band (i.e., the lower flat band) from graphene valley $\mathbf{K}_{\xi_{-}}^{\textrm{FLG}}$
obtained from the continuum model (a)-(d) and Wannier TB model (e)-(h).
The high symmetric points in the moir\'{e} BZ are presented. The parameters
of h-BN substrate $\Delta_{\textrm{BN}}^{d}$ and displacement field
$U_{\textrm{D}}$ are given in the main text. The interpolated Wannier
TB model can well describe the Berry curvature distributions, so does
the Chern number.}
\end{figure*}

In this section, we compare the Berry curvature distributions calculated
from the continuum model and Wannier TB model. The Berry curvature
for the n\textit{th} band can be calculated from\citep{prl1982_Thouless,prl2004_Yao}

\begin{eqnarray}
\Omega_{n}^{z}\left(\mathbf{k}\right) & = & -2\mathrm{Im}\,\sum_{m\neq n}\frac{v_{mn}^{x}\left(\mathbf{k}\right)v_{nm}^{y}\left(\mathbf{k}\right)}{\left(\omega_{m\mathbf{k}}-\omega_{n\mathbf{k}}\right)^{2}},
\end{eqnarray}
where $\varepsilon_{n\mathbf{k}}=\hbar\omega_{n\mathbf{k}}$ and $\boldsymbol{v}\left(\mathbf{k}\right)$
is the velocity operator matrix. And the Chern number is followed
by integration over the two-dimensional BZ

\begin{eqnarray}
\mathcal{C}_{n} & = & \frac{1}{2\pi}\int d^{2}k\,\Omega_{n}^{z}\left(\mathbf{k}\right).\label{eq:s.chern_sum_bc}
\end{eqnarray}
For Wannier TB model, the velocity operator $\boldsymbol{v}\left(\mathbf{k}\right)$
can be calculated by Wannier interpolation method\citep{prb2006_Wangxinjie_AHC}.
It should be notice that due to the extended shape of WFs for TMG
systems, one should include more matrix elements $\left\langle n\mathbf{0}|\hat{\mathbf{r}}|m\mathbf{R}\right\rangle ${[}see
Eq.(39) in Ref.\citep{prb2006_Wangxinjie_AHC}{]} to precisely describe
the velocity operator for Wannier TB model. It can be obtained by
numerically integrating the WFs in real space. We found however, by
including only the Wannier center $\left\langle n\mathbf{0}|\hat{\mathbf{r}}|n\mathbf{0}\right\rangle $,
one can get reasonable results in Berry curvature calculations. For
the continuum model 

\begin{eqnarray}
v_{mn}^{a}\left(\mathbf{k}\right) & = & \sum_{XX^{\prime}}\sum_{\mathbf{G}\mathbf{G}^{\prime}}C_{m\mathbf{k}}^{X}\left(\mathbf{G}\right)^{\ast}\left\langle \tilde{\psi}_{\mathbf{k}_{\xi}^{d}+\mathbf{G}}^{X}|i\hbar^{-1}\left[\hat{H},\hat{r}^{a}\right]|\tilde{\psi}_{\mathbf{k}_{\xi^{\prime}}^{d^{\prime}}+\mathbf{G}^{\prime}}^{X^{\prime}}\right\rangle C_{n\mathbf{k}}^{X^{\prime}}\left(\mathbf{G}^{\prime}\right)\nonumber \\
 & = & \sum_{XX^{\prime}}\sum_{\mathbf{G}\mathbf{G}^{\prime}}C_{m\mathbf{k}}^{X}\left(\mathbf{G}\right)^{\ast}\left\langle \tilde{u}_{\mathbf{k}_{\xi}^{d}+\mathbf{G}}^{X}|\frac{\partial\hat{H}\left(\mathbf{k}\right)}{\hbar\partial k_{a}}|\tilde{u}_{\mathbf{k}_{\xi^{\prime}}^{d^{\prime}}+\mathbf{G}^{\prime}}^{X^{\prime}}\right\rangle C_{n\mathbf{k}}^{X^{\prime}}\left(\mathbf{G}^{\prime}\right),
\end{eqnarray}
where

\begin{eqnarray}
\left\langle \tilde{u}_{\mathbf{k}_{\xi}^{d}+\mathbf{G}}^{X}|\frac{\partial\hat{H}\left(\mathbf{k}\right)}{\hbar\partial k_{a}}|\tilde{u}_{\mathbf{k}_{\xi^{\prime}}^{d^{\prime}}+\mathbf{G}^{\prime}}^{X^{\prime}}\right\rangle  & = & \delta_{\xi\xi^{\prime}}\delta_{\mathrm{sign}\left(d\right),\mathrm{sign}\left(d^{\prime}\right)}\delta_{\mathbf{G}\mathbf{G}^{\prime}}\nonumber \\
 &  & \times i\sum_{\mathbf{R}}\left(R^{a}+\tau_{\beta}^{a}-\tau_{\alpha}^{a}\right)e^{i\left(\mathbf{k}_{\xi}^{d}+\mathbf{G}\right)\cdot\left(\mathbf{R}+\tau_{\beta}-\tau_{\alpha}\right)}\left\langle \tau_{\alpha},d,\mathbf{0}+\mathbf{d}|\hat{H}_{0}|\tau_{\beta},d^{\prime},\mathbf{R}+\mathbf{d}^{\prime}\right\rangle .
\end{eqnarray}

\begin{table*}
\begin{centering}
\caption{\label{TABLE.s.chern}Chern number for twisted multilayer graphenes.
It is obtained by integrating the Berry curvature over the BZ {[}see
Eq.(\ref{eq:s.chern_sum_bc}){]}. The same results are obtained from
Wilson loop method and efficient lattice method\citep{arxiv2005-Chern-latticeMethod}.
Here $\xi_{+}$ and $\xi_{-}$ label $\left(\mathbf{K},\mathbf{K}^{\prime}\right)$
valley, and VB, CB indicate the lower and higher flat bands respectively. }
\par\end{centering}
\begin{ruledtabular}
\begin{centering}
\begin{tabular}{ccr@{\extracolsep{0pt}.}lr@{\extracolsep{0pt}.}lr@{\extracolsep{0pt}.}lr@{\extracolsep{0pt}.}lr@{\extracolsep{0pt}.}l}
 & Flat band index & \multicolumn{2}{c}{mTBG($1.08^{\circ}$)} & \multicolumn{2}{c}{TDBG($2.0^{\circ}$)} & \multicolumn{2}{c}{mTDBG($1.248^{\circ}$)} & \multicolumn{2}{c}{mTTG($1.248^{\circ}$)} & \multicolumn{2}{c}{$\mathrm{TMG_{\{ABCA\}}^{\{ABCA\}}}$($2.646^{\circ}$)}\tabularnewline
\hline 
Continuum model & ($\xi_{-}$, VB) & $\quad$0&975 & $\quad$0&021 & $\quad\quad$-3&061 & $\quad\quad$-0&014 & $\quad$$\quad$$\quad$0&188\tabularnewline
 & ($\xi_{-}$, CB) & -0&996 & -0&031 & 3&027 & 0&954 & -0&182\tabularnewline
 & ($\xi_{+}$, VB) & -0&976 & -0&021 & 3&061 & 0&014 & -0&183\tabularnewline
 & ($\xi_{+}$, CB) & 0&996 & 0&031 & -3&027 & -0&954 & 0&176\tabularnewline
Wannier TB model & ($\xi_{-}$, VB) & 0&907 & 0&011 & -2&873 & 0&053 & 0&055\tabularnewline
 & ($\xi_{-}$, CB) & -0&981 & 0&004 & 2&929 & -0&127 & 0&008\tabularnewline
 & ($\xi_{+}$, VB) & -0&907 & -0&011 & 2&873 & -0&053 & -0&055\tabularnewline
 & ($\xi_{+}$, CB) & 0&981 & -0&004 & -2&929 & 0&127 & -0&008\tabularnewline
\end{tabular}
\par\end{centering}
\end{ruledtabular}
\end{table*}

The Berry curvature distributions of TBG, TDBG and TMG for both the
continuum model and Wannier TB model are shown in Fig.\ref{FIG.S6.bcdist}.
It can be well described by the interpolated Wannier TB model. The
numerical integration of the Berry curvature are presented in TABLE.\ref{TABLE.s.chern}
giving the Chern number for the flat band. For TMG system with vanishing
valley Chern number (the two flat bands together for single valley
have zero Chern number), our model gives the correct topological classification
comparing with the original continuum model. For TMG system with non-vanishing
valley Chern number, for instance the mTTG in our presented results,
despite the well fitted band structure, our current model fails to
describe the topological feature for mTTG. 
\end{document}